\documentclass[11pt,a4paper]{article}
\usepackage{jheppub}

\title{Effects of temperature on thick branes and the fermion (quasi-)localization}
\author{Zhen-Hua Zhao,
        Yu-Xiao Liu\footnote{Corresponding author},
        Yong-Qiang Wang,
        Hai-Tao Li}
\affiliation[]{Institute of Theoretical Physics,
    Lanzhou University, Lanzhou 730000,
    People's Republic of China}
\emailAdd{zhaozhh09@lzu.edu.cn}
\emailAdd{liuyx@lzu.edu.cn}
\emailAdd{yqwang@lzu.edu.cn}
\emailAdd{liht07@lzu.edu.cn}

\abstract{Following Campos's work [Phys. Rev. Lett. \textbf{88}, 141602 (2002)],
we investigate  the effects of temperature on flat, de Sitter (dS),
and anti-de Sitter (AdS) thick branes in five-dimensional (5D) warped spacetime, and
on the fermion (quasi-)localization.
First, in the case of flat brane, when the critical
temperature reaches, the solution of the background scalar field and the warp
factor is not unique. So the thickness of the flat thick brane is uncertain at the critical value of the temperature parameter, which is found to be lower than the one in flat 5D spacetime.
The mass spectra of the fermion Kaluza-Klein (KK) modes are continuous,
and there is a series of fermion resonances. The number and lifetime
of the resonances are finite and increase with the temperature parameter,
but the mass of the resonances decreases with the temperature parameter.
Second, in the case of dS brane, we do not find such a critical value of
the temperature parameter. The mass spectra of the fermion KK modes are also continuous,
and there is a series of fermion resonances. The effects of temperature on resonance
number, lifetime, and mass are the same with the case of flat brane.
Last,
in the case of AdS brane, {the critical value of the temperature parameter can less or greater
than the one in the flat 5D spacetime.}
The spectra of fermion KK modes are discrete,
and the mass of fermion KK modes does not decrease monotonically
with increasing temperature parameter. }
\keywords{Large Extra Dimensions, Field Theories in Higher Dimensions}
\begin{document}
\maketitle

\section{Introduction}

In the braneworld scenario our universe is a 3-brane embedded
in a higher dimensional spacetime.
The braneworld scenario has received a lot of attentions since it can
provide us a novel approach
to resolve the cosmological constant and the hierarchy problems \cite{Arkani-Hamed1998429,AntoniadisDimopoulosDvali1998,AntoniadisArkani-HamedDimopoulosDvali1998,Randall199983},
and reproduce the Newtonian law of gravity \cite{Randall199983a}.
In the braneworld scenario all kinds of
matter fields should be localized on the brane, and there is much literature on these issues
\cite{Bajc2000,Randjbar-Daemi2000,Pomarol2000,CasadioGruppusoVenturi2000,CasadioGruppuso2001,Akhmedov2001,
Dvali2001,Ghoroku2002,Ringeval200265,
Koley200522,Oda2003,Melfo2006,Liu2007a,Liu2007,Davies2007,Slatyer2007,KoleyMitraSenGupta2010,
Liu200878,Liu200808,Liu200802,Guerrero2009,Zhao2009,
Liu2009a,Liu2010,CorreaDutraHott2010,Castro2011,CastroMeza2010}. The braneworld scenario also provides some new approaches to
resolve the family and favor problems in particle physics
\cite{FrereLibanovTroitsky2001,LibanovTroitsky2001,Neronov2002,
AguilarSingleton2006,GogberashviliMidodashviliSingleton2007,
GuoMa2008,GuoMa2009}.

Further, taking into account that our real universe is a system with the temperature.  So it is
interesting to  investigate the effects
of temperature in braneworld theories.
Brevik et al. \cite{Brevik2001599} investigated the quantum (in)stability of
the AdS$_5$ braneworld universe at nonzero temperature. Campos discussed the critical phenomena of flat
thick branes in warped spacetimes~\cite{Campos200288} with a five-dimensional (5D) complex background scalar field.
Bazeia et al. \cite{Bazeia200411}  investigated the geometric transitions of thick braneworlds
at high temperature limit.
Different from the ideal thin braneworld models \cite{Randall199983,Randall199983a},
thick braneworld models usually
need to introduce 5D background scalar fields to generate the branes
\cite{Bonjour1999,Goldberger1999,Csaki2000581,DeWolfe200062,Gremm2000478,Ichinose200118,
Ichinose200118a,Kehagias2001504,Campos200288,Gregory2002,Gregory2002a,Kobayashi2002,
Bazeia200391,Bronnikov2003,Eto200368,Melfo2003,Bazeia2004,Bazeia200405,Bazeia2006f,
Dzhunushaliev2006,Dzhunushaliev200713,George2007,Dzhunushaliev200877,
Bazeia2009c,Burnier2009,Chumbes2010,Dzhunushaliev200979,Liu2009c}.
To introduce the temperature effects to thick braneworld theories,
we can calculate the effective potential of the 5D background scalar fields at finite temperature.
And the method is the same as the one proposed in \cite{Dolan1974,Jackiw1975,Weinberg1974}
in four-dimensional flat spacetime.
Ansari and Suresh \cite{Ansari200722} calculated such an effective potential of $\phi^4$
model in 5D flat spacetime with the one-loop
correction. As far as we know, the effective potential of 5D scalar fields
in thick braneworld theories has not been calculated. To do such a calculation is very meaningful
but also difficult, not only because the
spacetime is curved, but also because the background scalar field is
coupled with the gravity. Although without
an analysis function form of the effective potential with the temperature,
one can still investigate the
temperature effects in braneworld theories
\cite{Campos200288,Bazeia200411}. Within the $\phi^4$ model, only the mass parameter
is corrected by the effects of temperature,
so the variation of the mass parameter will reflect the effects of temperature on branes.

Our study is based on the work of Campos~\cite{Campos200288}. In this paper, the self-interaction potential of the background complex scalar field $\Phi$ has the form of ~\cite{Campos200288}
\begin{equation}
V(\Phi)=a|\Phi|^2-b\phi_{\rm R}(\phi_{\rm R}^2-3\phi_{\rm I}^2)+c|\Phi|^4+C_5,\label{V(Phi)}
\end{equation}
where $\phi_{\rm R}$ and $\phi_{\rm I}$ are the real and imaginary parts of the complex scalar
field $\Phi$, respectively, and $C_5$ is a constant. If there is no background complex
scalar field $\Phi$,
$C_5$ will be the cosmological constant of 5D spacetime.
Taking the effects of temperature into account, the mass parameter $a$
should be a function of the temperature $T$ \cite{Dolan1974,Weinberg1974,Jackiw1975}.
But the parameters $b$ and $c$ will not vary with the temperature.
This can be seen from $V(\Phi)$ after a shift of $\Phi$
around its vacuum expectation value $\Phi_0(\phi_{\rm R 0}, \phi_{\rm I0})$:
$\Phi\rightarrow \delta \Phi+\Phi_{\rm R}$ [$\phi_{\rm R}\rightarrow\delta\phi_{\rm R}+\phi_{\rm R 0}$,
$\phi_{\rm I}\rightarrow\delta\phi_{\rm I}+\phi_{\rm I 0}$]. After that shift the coefficient before
$|\Phi_0|^4$ is also $c$, so $c$ will not vary with the temperature.
The coefficients before $\phi_{\rm R0}^3$ and $\phi_{\rm R0}\phi_{\rm I0}^2$ are
changed from $-b$ and $3b$ to $-b+2c\delta\phi_{\rm R}$ and $3b+4c\delta\phi_{\rm R}$, respectively.
Because the vacuum expectation value of $\delta\phi$: $\langle 0|\delta\phi_{\rm R}|0\rangle=0$, so the coefficients
$b$ will also not vary with the temperature.
Because we have not the exact function form of $a(T)$, we only take $a$ as the temperature parameter in this paper.

Campos found that the presence of gravity would lower the critical value of the
temperature parameter of the phase
transition comparing to 5D flat spacetime~\cite{Campos200288}.
And the phase transition is characterized by the emergence of a double kink
solution of the $\phi_{\rm I}$.  Below the critical
value of the temperature parameter, the solution of the $\phi_{\rm I}$ has a single kink form.
Further, we find that, in the case of AdS brane, the critical value of $a$  {can less or greater
than the one in the 5D flat spacetime,}
and we do not find the double kink solution of the $\phi_{\rm I}$ for the case of dS brane.

We also investigate the effects of temperature on fermion localization
and quasi-localization on flat, dS, and AdS thick Branes.
In order to localize fermions on thick branes,
the coupling between fermions and the background scalar fields is needed.
We consider the Yukawa coupling  $\eta\bar\Psi\phi_{\rm I}\Psi$.
We find that, in the cases of flat and dS branes, the fermion zero mode can be localized on
the branes, and there are quasi-localized massive fermion KK modes, namely, fermion resonances.
The number and the lifetime of the fermion resonances
increase with the temperature parameter $a$.
But the resonance mass decreases with the increase of $a$.
In the case of AdS brane,
the spectrum of fermion KK modes is discrete,
and the number of the discrete KK modes is infinite.
The variation of masses of fermion KK modes does not decrease monotonically with the increase
of $a$.

The paper is organized as follows.
Firstly, the effects of temperature on
flat, dS, and AdS thick branes are discussed in section~\ref{sec:BRANES}.
Secondly, the effects of temperature on the fermion localization and fermion resonances are investigated in section~\ref{sec:Fermions}.
For self-completeness and self-consistency, the Schr\"{o}dinger equations and the corresponding potentials for the KK modes of left- and right-handed fermions are derived.
Before performing numerical calculations, we analyze the asymptotic behaviors of the potentials at zero and infinity along the extra dimension.
In sbusection~\ref{sec:flatds}, we investigate the effects of temperature on
the number, lifetime, and mass of fermion resonances quasi-localized on flat and dS branes.
We show the variation of masses of fermion KK modes localized on AdS brane
with the temperature parameter $a$ in subsection~\ref{sec:ads}.
Finally, the conclusion and discussion are given in section~\ref{sec:conclusion}.

\section{Effects of temperature on thick branes} \label{sec:BRANES}

To investigate  the  effects  of temperature  on {thick} branes,  we consider a 5D action with {a complex scalar field} $\Phi$ minimally coupled to gravity:
\begin{equation}
       S = \int d^4x dy \sqrt{-g}
           \left(  \frac{1}{4} R
             - \frac{1}{2} g^{MN} {\partial_M \Phi^{*}\partial_N \Phi}
             - V(\Phi)
           \right),\label{action2}
\end{equation}
where $V(\Phi)$ is the same  as the one in Eq. (\ref{V(Phi)}). To ensure the stability{,} the parameter $c$ must be positive. There are three degenerate global minima for the potential $V(\Phi)$ if $b>0$ and $0<a<9a_c/8$, where $a_c=b^2/4c$~\cite{Campos200288}. The three degenerate global minima {are at} $\Phi^1=\phi_0$, $\Phi^2=(-\frac{1}{2}+i\frac{\sqrt{3}}{2})\phi_0$, and $\Phi^3=(-\frac{1}{2}-i\frac{\sqrt{3}}{2})\phi_0$, where $\phi_0=\frac{3b}{8c}(1+\sqrt{1-\frac{8a}{9a_{c}}})$~\cite{Campos200288}.

The line element in this model is assumed as
\begin{eqnarray}
  ds^2=e^{2A(z)}(\hat g_{\mu\nu}(x)dx^\mu dx^\nu+dz^2),\label{metric}
\end{eqnarray}
where $\mu, \nu=0,1,2,3$, $\hat g_{\mu\nu}$ is the 4-dimensional metric on branes, and $e^{2A(z)}$ is the warp factor. The usual hypothesis is that $A$ {and $\Phi$} are only the functions of the extra dimension coordinate $z$.

{For the three types of maximally 4-symmetric branes (flat, dS and AdS)}, {the equations of motion of $A(z)$, $\phi_{\rm I}$, and $\phi_{\rm R}$ have} {a unified} form:
\begin{eqnarray}
  \phi_{\rm I}''&=&-3 A'\phi_{\rm I}'+e^{2A}\frac{dV}{d\phi_{\rm I}}\label{eq_cf_phiI},\\
  \phi_{\rm R}''&=&-3 A'\phi_{\rm R}'+e^{2A}\frac{dV}{d\phi_{\rm R}}\label{eq_cf_phiR},\\
  A''&=&A'^2 -\frac{1}{3}((\phi_{\rm I}'^2+\phi_{\rm R}'^2)+\Lambda_4)\label{eq_cf_A''},\\
  6A'^2&=&2\Lambda_4+\frac{1}{2}\big( (\phi_{\rm I}'^2+\phi_{\rm R}'^2)-2 e^{2 A}V\big)\label{C_5_cf},
\end{eqnarray}
where the prime stands for the derivative with respect to $z$.
$\Lambda_4$ is the cosmological constant on branes, and ${\Lambda_4}=0$, $>0$ and $<0$ {correspond to the cases of flat, dS and AdS branes}, respectively. Equations (\ref{eq_cf_phiI})-(\ref{C_5_cf})
are a set of nonlinear ordinary differential equations, which can be
solved numerically
\footnote{We can solve these equations with the FORTRAN code colsys and bvp\_solver,
or the function bvp5c in Matlab.}
with the following boundary conditions:
\begin{eqnarray}
A(0)=A'(0)=\phi_{\mathrm R}'(0)=\phi_{\mathrm I}(0)=0,\\
\phi_{\rm R}(+\infty)=-\frac{1}{2}\phi_0,\;\; \rm{and}\;\;\phi_{\rm I}(+\infty)=\frac{\sqrt{3}}{2}\phi_0.\label{28}
\end{eqnarray}

\begin{figure*}
  \begin{center}
  \includegraphics[width=0.45\textwidth]{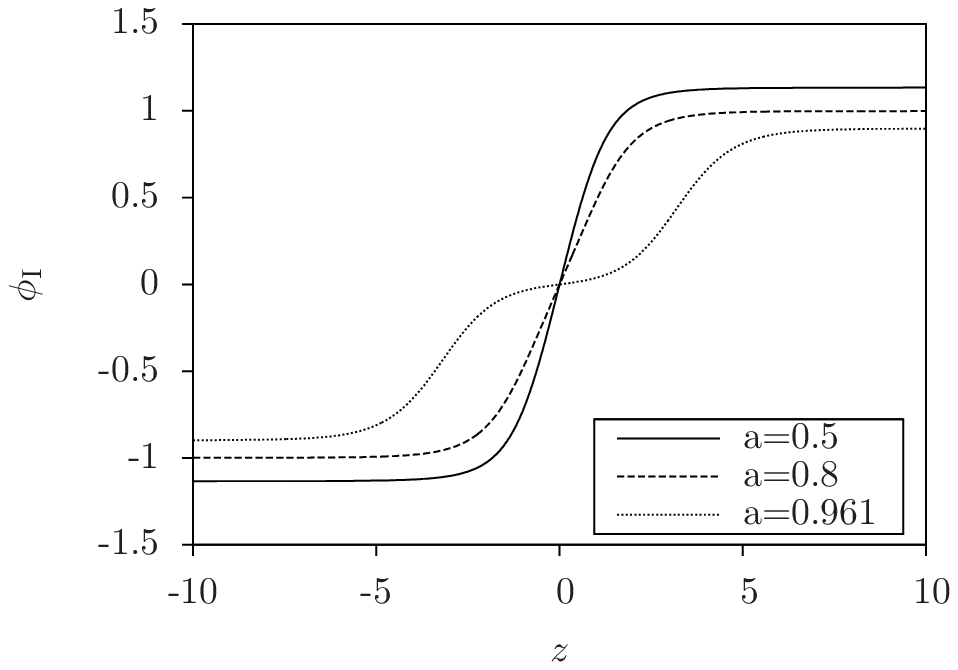}
  \includegraphics[width=0.45\textwidth]{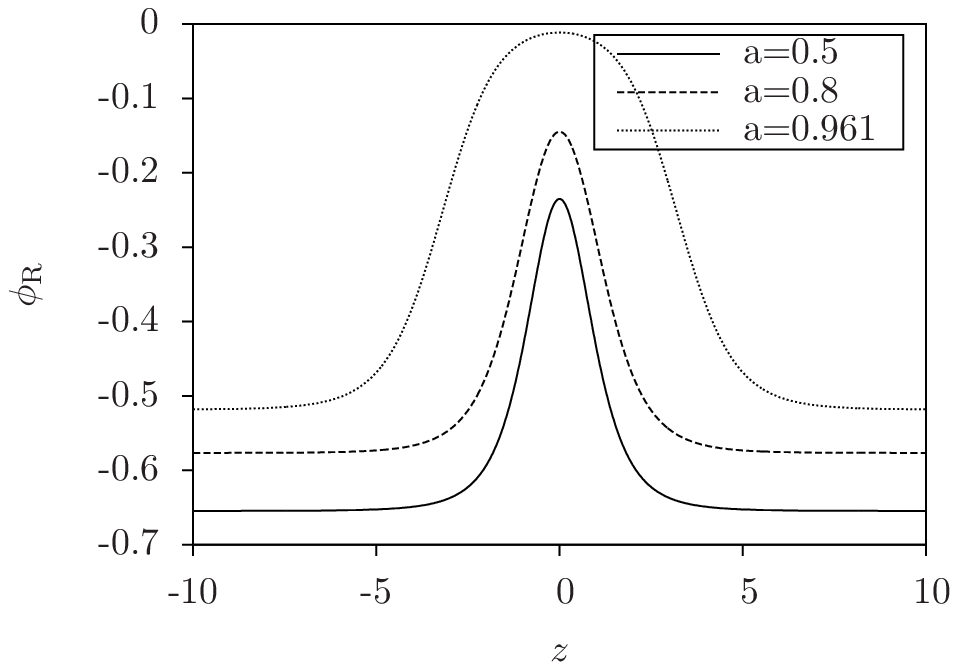}\\
\caption{The profiles of the background scalar fields $\phi_{\mathrm I}$ and $\phi_{\mathrm R}$
for flat brane. The three lines correspond to $a=0.5, 0.8, 0.961$, respectively.
{The other parameters} are set to $b=2$ and $c=1$.}\label{fig_cf_flat_Phi}
  \end{center}
\end{figure*}

\begin{figure}
  \begin{center}
  \includegraphics[width=0.45\textwidth]{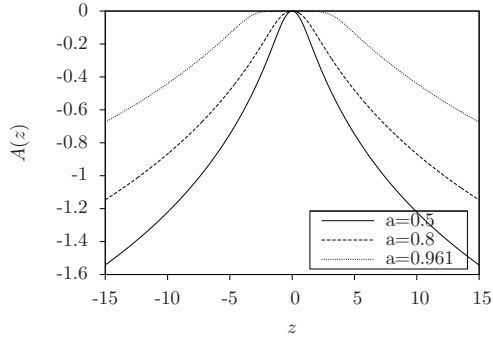}\\
    \caption{The profiles of the warp factor $A$ for flat brane. The three lines correspond to $a=0.5, 0.8, 0.961$, respectively. The other parameters are set to $b=2$ and $c=1$.}\label{fig:FactorA_flat}
  \end{center}
\end{figure}

The solutions of $A$, $\phi_{\mathrm I}$, and $\phi_{\mathrm R}$
are varied with the temperature parameter $a$. For flat, dS, and AdS
thick branes the variation of the
solutions of $\phi_{\mathrm I}$ and $\phi_{\mathrm R}$ with $a$ is similar.
So, we only show the solutions for the case of flat brane in Fig.~\ref{fig_cf_flat_Phi}  with $b=2$, $c=1$, and different values of $a$.
Fig.~\ref{fig_cf_flat_Phi} shows that the solutions of flat thick brane are dilated along the
extra dimension with the increase of $a$, and the solution of $\phi_{\mathrm I}$ tends to have a double kink form.

The solutions of $A(z)$ for flat and dS branes are similar,
and the solutions for the case of flat brane are shown in Fig.~\ref{fig:FactorA_flat}.

\begin{figure}
\begin{center}
\includegraphics[width=0.45\textwidth]{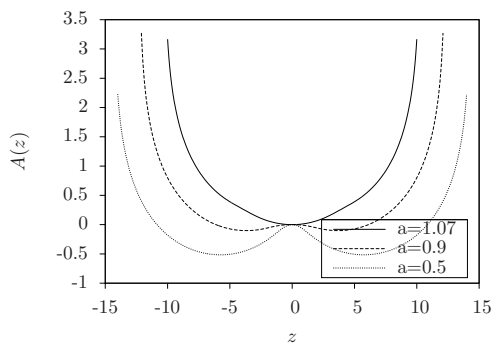}
\end{center}
\caption{ \label{fig:A_ads}
The profiles of {the warp factor $A(z)$ for AdS brane. The three lines correspond to  $a=1.07$, $0.9$, and $0.5$.} The other parameters are set to
 $b=2$, $c=1$, and $\Lambda_4=-0.1$.}
\end{figure}

For the case of AdS brane, the behavior of the warp factor is very different
from the cases of flat and dS branes. This has been analyzed in Refs. \cite{KarchRandall2001,Gremm2000,Liu2010}.
In this case, with the metric (\ref{metric}) and the potential (\ref{V(Phi)}), the $z$ coordinate only
runs from $-z_b$ to $z_b$, where $z_b$ is a finite value, and $A(z)$ is
divergent at {${\pm}z_b$}. This can be {seen} from the analysis of the solutions of Eqs.~(\ref{eq_cf_phiI})-(\ref{C_5_cf}) at $z\rightarrow\pm z_b$.
When $z\rightarrow\pm z_b$,
\begin{eqnarray}
\phi_{\rm R}(\pm z_b)=-\frac{1}{2}\phi_0,\;\; \phi_{\rm I}(\pm z_b)=\pm\frac{\sqrt{3}}{2}\phi_0,\label{29}
\end{eqnarray}
and
\begin{eqnarray}
\phi_{\mathrm R}'(\pm z_b)=\phi_{\mathrm I}'(\pm z_b)\rightarrow0.
\end{eqnarray}
So Eqs.~(\ref{eq_cf_A''}) and (\ref{C_5_cf}) turn into
\begin{eqnarray}
  A''(z\rightarrow\pm z_b)&=&A'^2(z\rightarrow\pm z_b) -\frac{1}{3}\Lambda_4 \label{eq_ads_A''},\\
  6A'^2(z\rightarrow\pm z_b)&=&2\Lambda_4- e^{2 A(z\rightarrow\pm z_b) }{V_{z_b}} \label{C_5_ads},
\end{eqnarray}
where ${V_{z_b}}{\equiv}V(\phi_{\rm I}(\pm z_b),\phi_{\rm R}(\pm z_b))$ is a constant. The solution of the above equations is
\begin{eqnarray}
A(z\rightarrow\pm z_b)\rightarrow \log\sqrt{{\frac{{2\Lambda_4}}{V_{z_b}}}}
-\log\left[\sin\left(\sqrt{\frac{-\Lambda_4}{3}}(z_b-|z|)\right)\right]. \label{A_adsinf}
\end{eqnarray}
{So it is clear that the warp factor $A(z)$ is divergent at the boundaries $z={\pm}z_b$}.
This is consistent with the numerical solution of $A(z)$ {shown in Fig.~{\ref{fig:A_ads}}, from which it can be seen that the value of $z_b$ decreases with increasing $a$.}

The critical value of the bulk temperature parameter $a$ is defined such that
the profile of $\phi_{\mathrm I}$ has the form of a double kink \cite{Campos200288}, {namely $\phi'_{\rm I}(0)=0$}.
For the case of five-dimensional flat spacetime ($A=0$), we can easily determine the critical value of $a$, which is given by $a_c=b^2/(4c)$ \cite{Campos200288}.
But for {five-dimensional warped spacetime with flat brane},
the critical value of $a$ is not $a_c$ but a smaller effective critical value
$a_*$~\cite{Campos200288}.

\begin{figure*}
  \begin{center}
  \includegraphics[width=0.45\textwidth]{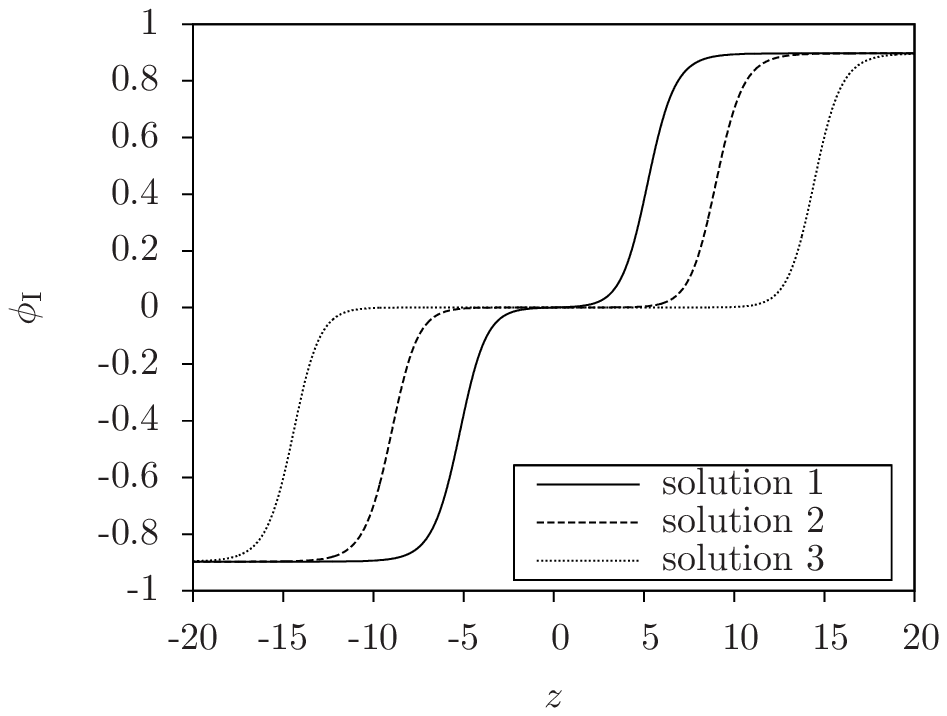}
  \includegraphics[width=0.45\textwidth]{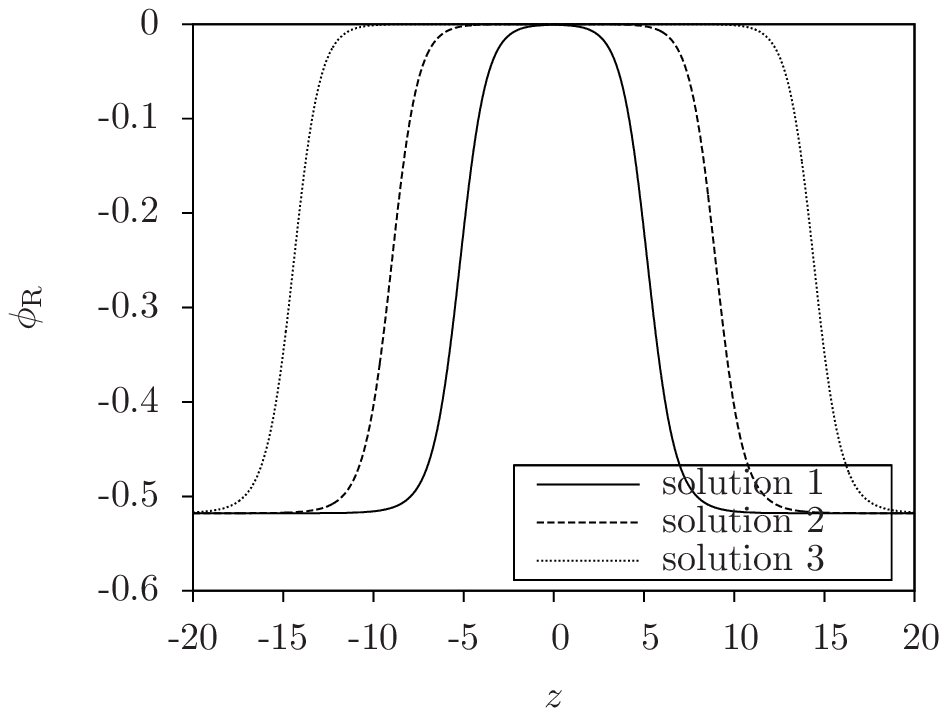}\\
    \caption{The profiles of $\phi_{\mathrm I}$ and $\phi_{\mathrm R}$
    with different solutions with $a_*=0.9617$ for flat branes.
    The other parameters are set to $b=2$ and  $c=1$.}\label{fig_cf_flat_Phi_critical}
  \end{center}
\end{figure*}

We find when $a=a_*$ the solutions of $\phi_{\mathrm I}$,
$\phi_{\mathrm R}$, and $A$ are not unique~\footnote{
Because Eqs. (\ref{eq_cf_phiI}-\ref{C_5_cf}) are non-linear, so in order to
solve them numerically we need to offer initial guess solutions of
$\phi_{\mathrm I}$, $\phi_{\mathrm R}$, and $A$.
And different guess solutions follow different solutions of them.}.
This means that at the critical temperature the width of flat thick brane is not fixed.
The profiles of $\phi_{\mathrm I}$ and $\phi_{\mathrm R}$ with $a_*=0.9617~(b=2,~c=1)$
are shown in Fig.~\ref{fig_cf_flat_Phi_critical} for flat brane.
In the AdS brane case, we find that the effective critical value $a_*$ {can} greater than $a_c$.
But for the dS brane, we do not find a double kink solution of $\phi_{\mathrm I}$.

In the Appendix we have discussed the relation between $a_*$ and $a_c$
with the assumptions that as $a\rightarrow a_*$,  $|(a_c-a_*)/a_c| \ll 1$ and $\phi_{\rm R}(0)^2\ll \phi_0^2$.
{And we show that, for the case of flat brane, $a_*<a_c$.
But in the cases of dS and AdS branes, the relation between $a_*$ and $a_c$ is uncertain.}

\begin{figure}
  \begin{center}
  \includegraphics[width=0.45\textwidth]{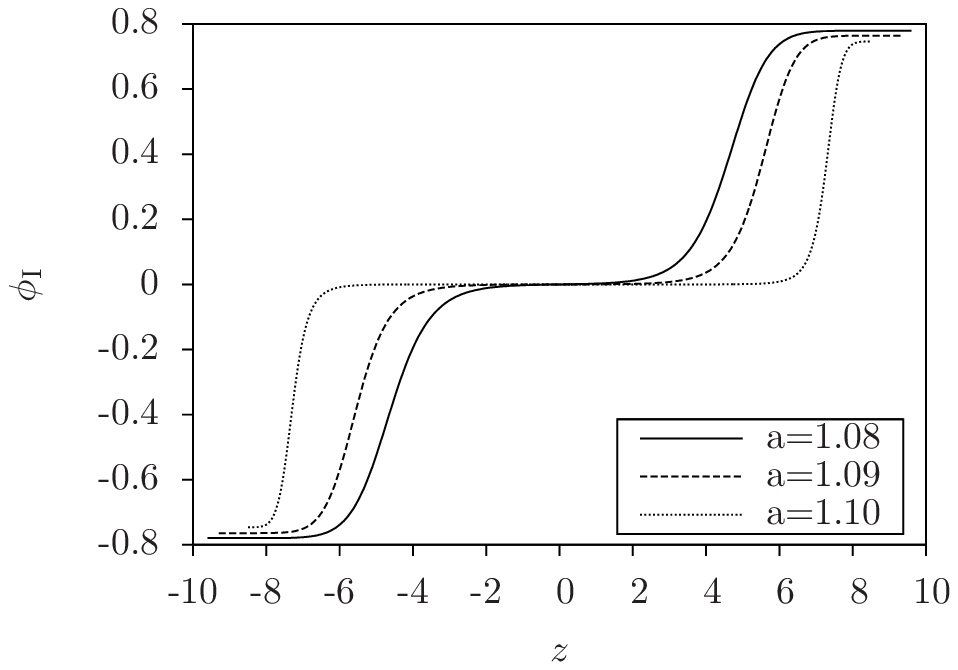}
   \includegraphics[width=0.45\textwidth]{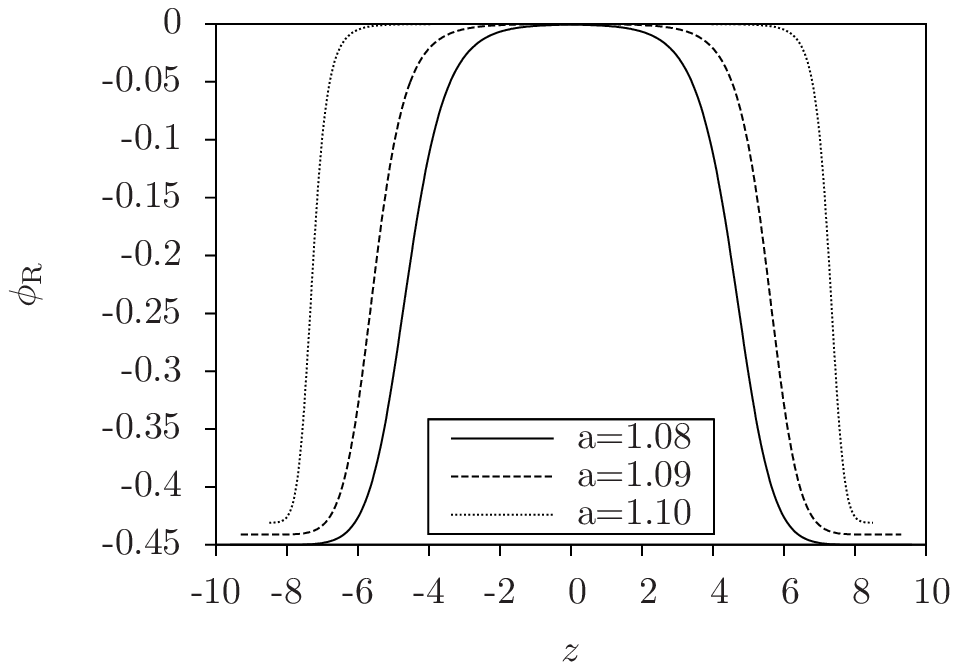}
    \caption{The profiles of $\phi_{\mathrm I}$ and $\phi_{\mathrm R}$ with $a=1.08$, $a=1.09$, and $a=1.10$
    for {AdS brane}.
     The critical value $a_{*}$ is {$1.084$}. The other parameters are set to $b=2$, $c=1$,
      and $\Lambda_4=-0.1$.}\label{fig:phiIupCritical_ads}
  \end{center}
\end{figure}

In the flat brane case, {when the value of $a$} is greater than the critical value $a_*$,
we do not find any solution of
Eqs.~(\ref{eq_cf_phiI})-(\ref{C_5_cf}). {However, in AdS brane case, we do,
and we show the profiles of
$\phi_{\mathrm I}$ {in Fig.~{\ref{fig:phiIupCritical_ads}}
for} $a=1.08$, $a=1.09$, and $a=1.10$ (the
critical value {$a_{*}$} is {$1.084$)}.}

\section{Effects of temperature on fermion localization and resonances}\label{sec:Fermions}

In order to localize a bulk fermion on these branes, we need to introduce the coupling of the fermion
and the background scalar $\Phi$. A general form of the coupling is $\eta\bar\Psi F(\phi_{\rm I},\phi_{\rm R})\Psi$, where $\Psi$ is the fermion field and $F(\phi_{\rm I},\phi_{\rm R})$ is a function of $\phi_{\rm I}$ and $\phi_{\rm R}$.
The action of the fermion describing such coupling is
\begin{eqnarray}
 S_{\Psi}=\int {d^5x}\sqrt{-g}\Big\{\bar\Psi\Gamma^M(\partial_M+\omega_M)\Psi-\eta\bar\Psi
 F(\phi_{\rm I},\phi_{\rm R})\Psi\Big\},
 \end{eqnarray}
where $\Gamma^M=(e^{-A}\gamma^{\mu},e^{-A}\gamma^{5})$ are the curved space gamma matrices,
$\eta$ is the coupling constant and $\eta>0$.
$\omega_M$ is the spin connection \cite{Duan1958,Fischbach1981, Zhao200776}
and its nonvanishing components are \cite{Liu2009f, Ringeval200265}
\[
\omega_{\mu}=\frac{1}{2}A'\gamma_{\mu}\gamma_{5}+\hat\omega_{\mu},
\]
where $\hat\omega_{\mu}$ is the spin connection derived from the metric $\hat g_{\mu\nu}$.
The resulting Dirac equation is
\begin{eqnarray}
 \Big[\gamma^{\mu}(\partial_{\mu}+\hat\omega_{\mu})+\gamma^5(\partial_z+2 A')-\eta
e^{A}F(\phi_{\rm I},\phi_{\rm R})\Big]\Psi=0,
\end{eqnarray}
where $\gamma^{\mu}(\partial_{\mu}+\hat\omega_{\mu})$ is the Dirac operator on branes.
With the usual KK and chiral decomposition
\begin{eqnarray}
 \Psi(x,z)=e^{-2A}\sum_n \Big(\psi_{{\rm{L}} n}(x)f_{{\rm{L}} n}(z)
 +\psi_{{\rm{R}} n}(x)f_{{\rm{R}} n}(z)\Big),
\end{eqnarray}
where subscripts `$\rm{L}$' and `$\rm{R}$' denote the Left- and right-handed {chiralities}, respectively,
and the {four-dimensional} massive Dirac equation
\begin{eqnarray}
\gamma^{\mu}(\partial_{\mu}+\hat\omega_{\mu})\psi_{{\rm{L}} n}&=&m_n\psi_{{\rm{R}} n},\\
\gamma^{\mu}(\partial_{\mu}+\hat\omega_{\mu})\psi_{{\rm{R}} n}&=&m_n \psi_{{\rm{L}} n},
\end{eqnarray}
we can obtain the coupled equations of $f_{{\rm{L}} n}$ and $f_{{\rm{R}} n}$
\cite{Liu200878}:
\begin{eqnarray}
\big[\partial_{z}+\eta e^A
F(\phi_{\rm I},\phi_{\rm R})\big]f_{{\rm{L}} n}(z)&=&\;\;m_n f_{{\rm{R}} n}(z),\label{eq:zeroMode}\\
\big[\partial_{z}-\eta e^A
F(\phi_{\rm I},\phi_{\rm R})\big]f_{{\rm{R}} n}(z)&=&-m_n f_{{\rm{L}} n}(z),~~~~~
\end{eqnarray}
where $f_{{\rm{L}} n}$ and $f_{{\rm{R}} n}$ satisfy the following orthonormality conditions:
\begin{eqnarray}\label{orthonormalityConditions}
 \int_{-\infty}^{\infty}f_{{\rm{L}} m}f_{{\rm{R}} n}dz
    =\delta_{m n}\delta_{{\rm{L}} {\rm{R}}}.
\end{eqnarray}
Further, we obtain the
Schr\"{o}dinger equations
 \cite{Liu200878}:
\begin{subequations}
\label{eq:motionF}
\begin{eqnarray}
  &&\big[-\partial_z^2+U_{\rm{fL}}(z)\big] f_{{\rm{L}} n}
    = m^2_n f_{{\rm{L}} n},\\
  &&\big[-\partial_z^2+U_{\rm{fR}}(z)\big] f_{{\rm{R}} n}
    = m^2_n f_{{\rm{R}} n},
\end{eqnarray}
\end{subequations}
where the effective potentials are given by
\begin{subequations}\label{eq:V}
\begin{eqnarray}
  U_{\rm{fL}}(z)&=& \eta^2 e^{2A}F(\phi_{\rm I},\phi_{\rm R})^2
                     -\eta e^A F(\phi_{\rm I},\phi_{\rm R})'
                    -\eta F(\phi_{\rm I},\phi_{\rm R}) e^A A',\\
  U_{\rm{fR}}(z)&=&U_{\rm{fL}}(z)|_{\eta\rightarrow-\eta}.
\end{eqnarray}
\end{subequations}

In order to determine the forms of $U_{\rm{fL}}(z)$ and $U_{\rm{fR}}(z)$, we should first {construct} the form of
$F(\phi_{\rm I},\phi_{\rm R})$. There are two simplest forms of $F$: $F_1=\phi_{\rm I}$ and $F_2=\phi_{\rm R}$.
With the solutions of $\phi_{\rm I}$ and $\phi_{\rm R}$,
we find that $\phi_{\rm I}$ is {an} odd function, and $\phi_{\rm R}$ is {an} even function.
So, in order to hold the $Z_2$ symmetry of $U_{\rm{fL}}(z)$ and $U_{\rm{fR}}(z)$,
we choose the first form of $F$, i.e., $F_1=\phi_{\rm I}$. So
\begin{subequations}
\begin{eqnarray}
  U_{\rm{fL}}(z)&=& \eta^2 e^{2A}\phi_{\rm I}^2-\eta e^A \phi_{\rm I}'
  -\eta \phi_{\rm I} e^A A', \label{Vf_L}\\
  U_{\rm{fR}}(z)&=& \eta^2 e^{2A}\phi_{\rm I}^2+\eta e^A \phi_{\rm I}'
  +\eta \phi_{\rm I} e^A A'. \label{Vf_R}
\end{eqnarray}
\end{subequations}

\subsection{Flat and dS branes}{\label{sec:flatds}}

Before solving Eq.~{(\ref{eq:motionF})} numerically, let us first {analyze the behavior of}
$U_{\rm{fL}}(z)$ and $U_{\rm{fR}}(z)$ at $z=0$ and $z\rightarrow\pm\infty$.

{First, at} $z=0$, with the boundary conditions $A(0)=A'(0)=\phi_{\mathrm I}(0)=0$, we get
\begin{subequations}
\begin{eqnarray}
  U_{\rm{fL}}(0)&=& -{\eta \phi_{\rm I}'},\\
  U_{\rm{fR}}(0)&=& +{\eta \phi_{\rm I}'}.
\end{eqnarray}
\end{subequations}
And from Fig.~\ref{fig_cf_flat_Phi}, we see that $\phi_{\rm I}'(0)>0$,
so $U_{\rm{fL}}(0)>0$ and $U_{\rm{fR}}(0)<0$ for positive coupling constant $\eta$.

{Second,} when $z\rightarrow\pm\infty$, $\phi_{\mathrm I}'(z)=0$. So we have
\begin{subequations}
\begin{eqnarray}
  U_{\rm{fL}}(z\rightarrow\pm\infty)&=& \eta^2 e^{2A}\phi_{\rm I}^2 -\eta \phi_{\rm I} e^A A', \\
  U_{\rm{fR}}(z\rightarrow\pm\infty)&=& \eta^2 e^{2A}\phi_{\rm I}^2 +\eta \phi_{\rm I} e^A A'.
\end{eqnarray}
\end{subequations}
And Eqs.~(\ref{eq_cf_A''}) and (\ref{C_5_cf}) turn into
\begin{eqnarray}
  A''&=&A'^2 -\frac{1}{3}\Lambda_4 \label{eq_A''},\\
  6A'^2&=&2\Lambda_4- e^{2 A }{V_{\infty}} \label{A'},
\end{eqnarray}
where {${V_{\infty}}{\equiv} V(\phi_{\rm I}(\pm\infty),\phi_{\rm R}(\pm\infty))$} is a constant,
and %
$\phi_{\rm R}(\pm\infty)=-\frac{1}{2}\phi_0$, $\phi_{\rm I}(\pm\infty)=\pm\frac{\sqrt{3}}{2}\phi_0$.

\begin{figure*}
  \begin{center}
  \includegraphics[width=0.45\textwidth]{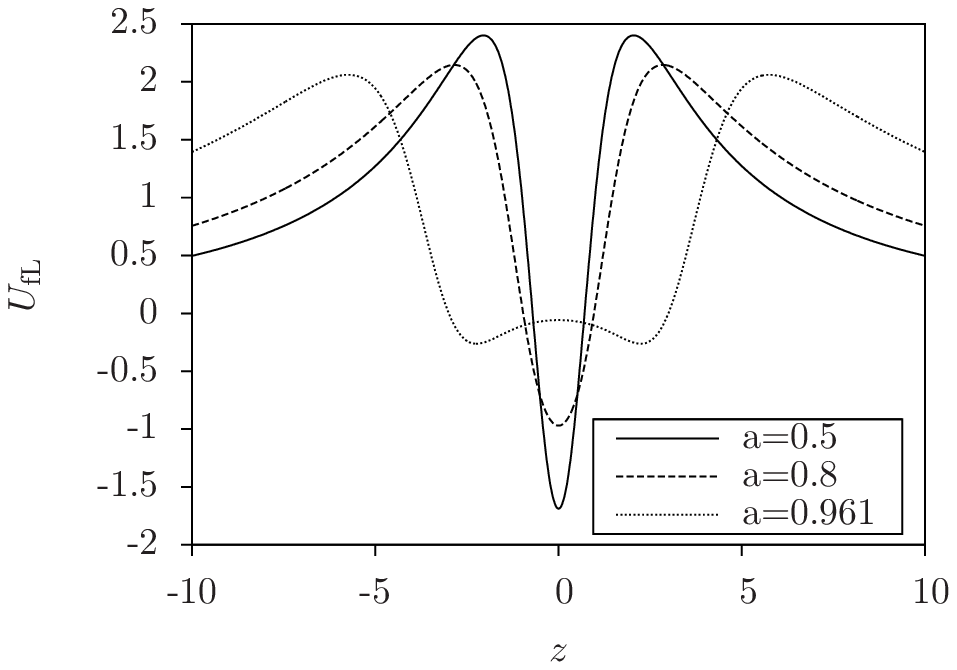}
  \includegraphics[width=0.45\textwidth]{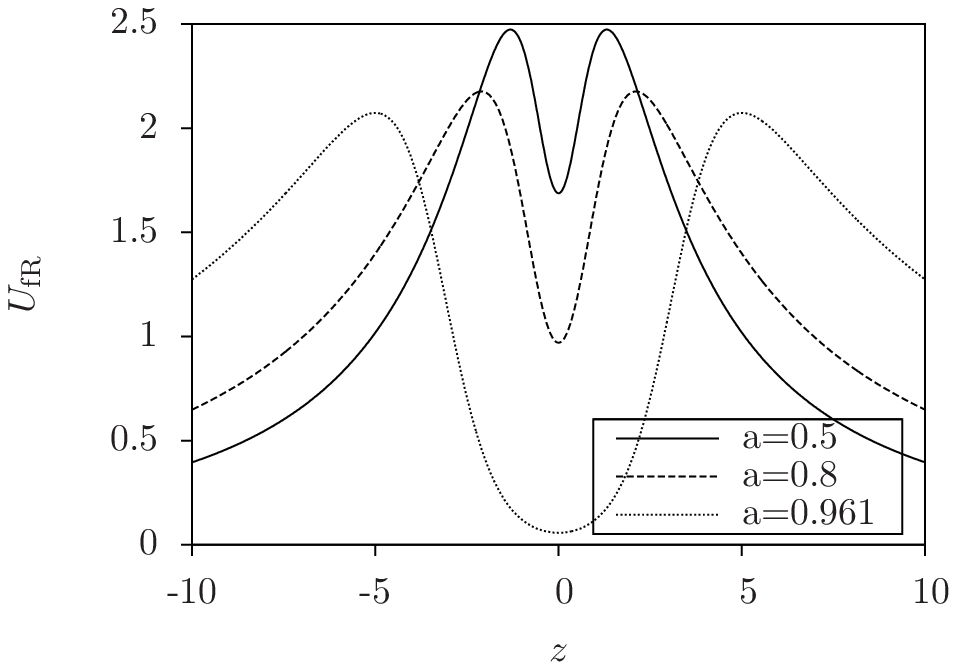}\\
    \caption{The profiles of the potentials of {fermion KK modes} for flat brane.
  {The parameters} are set to $\eta=2$, $b=2$, $c=1$, and {$a=0.5,0.8,0.961$}.}\label{fig:cf_flat_Ufermion}
  \end{center}
\end{figure*}

In the flat brane case, $\Lambda_4=0$, the solution of $A(z\rightarrow\pm\infty)$ is
\begin{eqnarray}
  A(z\rightarrow\pm\infty)=-\log\left({\sqrt{{-V_{\infty}}/{6}}}|z|+c_1\right),\label{Aflat}
\end{eqnarray}
{where  $c_1$ }is an integration constant.
So $e^{2A}|_{z\rightarrow\pm\infty}=1/({\sqrt{{-V_{\infty}}/{6}}}|z|+c_1)^2|_{z\rightarrow\pm\infty}\rightarrow 0$
and $A'(z)|_{z\rightarrow\pm\infty}=-1/({\sqrt{{-V_{\infty}}/{6}}}|z|+c_1)|_{z\rightarrow\pm\infty}\rightarrow 0$.
Then we reach the conclusion {that}
$U_{\rm{fR}}\rightarrow0$ and $U_{\rm{fL}}\rightarrow0$ when ${z\rightarrow\pm\infty}$.

In the dS  brane case, $\Lambda_4>0$, {the solutions of $A(z)$, $e^{A(z)}$, and $A'(z)$ at $z\rightarrow\pm\infty$ are}
\begin{eqnarray}
 A(z)&=& \log\bigg[\sqrt{\frac{2\Lambda_4}{{V_{\infty}}}}
        \text{sech}\left(\sqrt{\frac{\Lambda_4}{3}}(|z|+c_2)\right) \bigg],\label{A_dS} \\
 e^{A(z)} &=& \sqrt{\frac{2\Lambda_4}{{V_{\infty}}}}
        \mathrm{sech} \left(\sqrt{\frac{\Lambda_4}{3}}(|z|+c_2)
                      \right)\rightarrow 0,\\
 A'(z)&=&\mp\sqrt{\frac{\Lambda_4}{3}}
        \tanh\left(\sqrt{\frac{\Lambda_4}{3}}(|z|+c_2)\right)
        \rightarrow \mp\sqrt{\frac{\Lambda_4 }{3}}.
\end{eqnarray}
{Then we come to the conclusion that $U_{\rm{fL}}$ and $U_{\rm{fR}}$ vanish at ${z\rightarrow\pm\infty}$.}

\begin{figure}
\begin{center}
\includegraphics[width=0.45\textwidth]{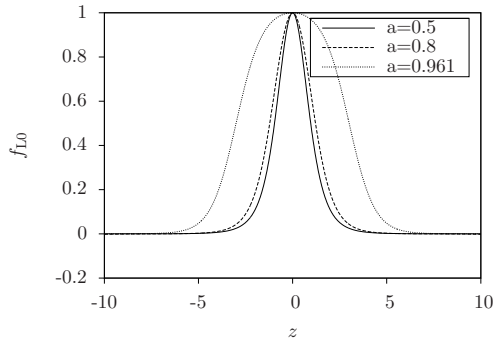}
\end{center}
\caption{ \label{fig:fermionZero_flat} The profiles of the zero mode of left-handed fermions for flat brane.  {The parameters} are set to $\eta=2$, $b=2$, $c=1$, and {$a=0.5,0.8,0.961$}.}
\end{figure}

With the numerical solutions of $A(z)$, $\phi_{\rm I}$, and $\phi_{\rm R}$,
we {plot} the potentials {$U_{\rm{fL}}$ and $U_{\rm{fR}}$} in
Fig.~{\ref{fig:cf_flat_Ufermion}} for flat brane. The profiles of {the} potentials for dS brane are {similar with the case of} flat brane,
and we {do not show them again}. We can see that, {for $\eta>0$, the potential of
left-handed KK modes $U_{\rm{fL}}$ is a modified volcano-type potential, which has a well lower than zero}, so only the left-handed fermions {have zero mode}. {The} Schr\"{o}dinger Eqs. (\ref{eq:motionF}) can be solved using the Numerov method \cite{GonzalezaThompsonb1997}
with two initial conditions at $z=0$, which can be set as
\begin{eqnarray}
 f_{{\rm{L}},{\rm{R}}}(0)=d_1,   f_{{\rm{L}},{\rm{R}}}'(0)=0\label{even}
\end{eqnarray}
for the even parity {KK modes} and
\begin{eqnarray}
  f_{{\rm{L}},{\rm{R}}}(0)=0,   f_{{\rm{L}},{\rm{R}}}'(0)=d_2\label{odd}
\end{eqnarray}
for the odd parity KK modes, where $d_1$ and $d_2$ are arbitrary constants. Here we choose $d_1=d_2=1$.
The profiles of the
zero mode for flat brane are shown in Fig.~{\ref{fig:fermionZero_flat}} with different values of
{the temperature parameter}. {The case for dS brane are similar}.

The volcano-type potential implies that there will not exist discrete spectrum of
fermion KK modes, but there {may} exist fermion resonances
~{\cite{Almeida2009,Liu2009f,Liu2009, Cruz2009}}.
The massive KK modes  cannot be normalized
because their wave functions are oscillating when far away from the brane along the extra dimension.
So Ref.~{\cite{Liu2009}} proposed a function
\begin{eqnarray}
P_{{\rm{L}},{\rm{R}}}(m)=
\frac{\int_{-z_c}^{z_c}|f_{{\rm{L}},{\rm{R}}}(z)|^2 dz}
     {\int_{-z_{max}}^{z_{max}}|f_{{\rm{L}},{\rm{R}}}(z)|^2 dz}
\end{eqnarray}
as the relative probability for finding the resonances on {a brane}.
{Here} $2z_c$ is about the width of the thick brane and $z_{max}$ is set to $z_{max}=10z_c$.
So for KK modes with $m^2\gg V_{{\rm{L}},{\rm{R}}}^{\text{max}}$ {(}$V_{{\rm{L}},{\rm{R}}}^{\text{max}}$
is the {maximum} value of $V_{{\rm{L}},{\rm{R}}}${)}, $f_{{\rm{L}},{\rm{R}}}$
can be approximately taken as plane waves, and the value of $P_{{\rm{L}},{\rm{R}}}(m)$
will {trend to} $0.1$.

\begin{figure*}
\includegraphics[width=0.45\textwidth]{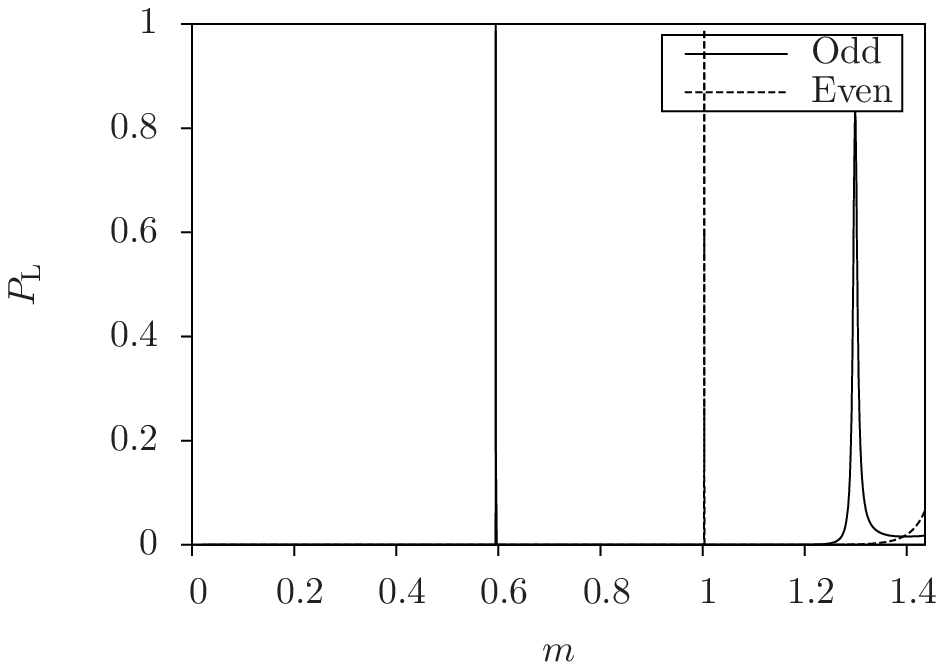}
\includegraphics[width=0.45\textwidth]{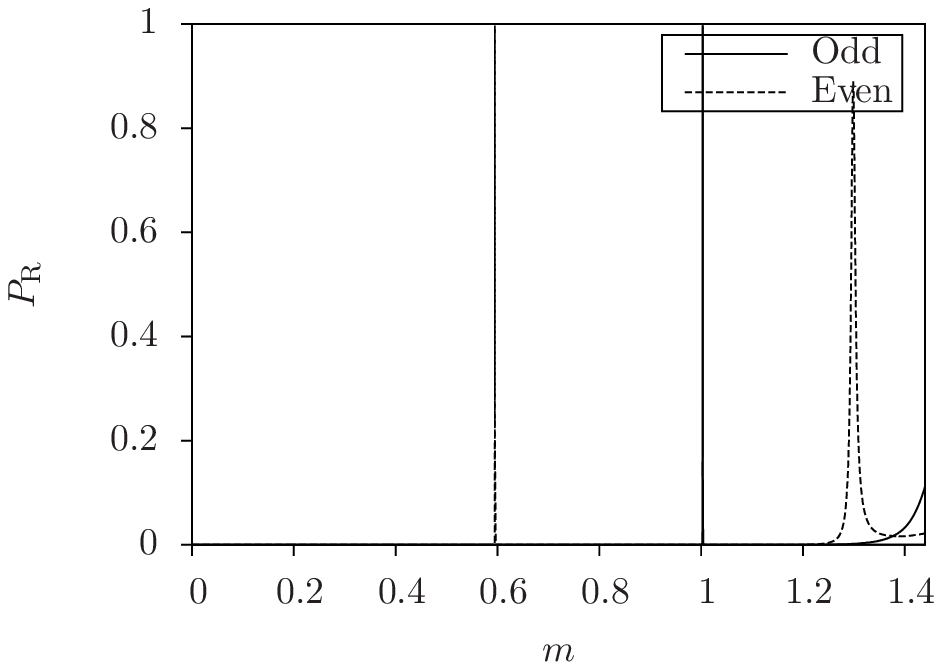}
\caption{ \label{fig:0961Spectra_flat}The profiles of $P_{{\rm{L}},{\rm{R}}}$ for left-handed fermions for flat brane with $a=0.961$, $b=2$, $c=1$, and $\eta=2$.}
\end{figure*}

As an example, we {plot} the profiles of $P_{{\rm{L}},{\rm{R}}}(m)$ in  Fig. {\ref{fig:0961Spectra_flat}}
corresponding to $a=0.961$ for flat brane.
In this figure each peak corresponds to a resonant state.
We can estimate the lifetime $\tau$ of {a resonance} as $\tau\sim \Gamma^{-1}$,
where $\Gamma=\delta m$ is the full width at half maximum of the peak \cite{GregoryRubakovSibiryakov2000a}.
It can be seen that {the spectra of resonances are} the same for both the
left- and right-handed fermions.

In order to investigate the effects of temperature on resonances.
We {solve} Eq.~{(\ref{eq:motionF})} with the following values of a:
\begin{displaymath}
\begin{array}{ll}
  a= \{ 0, 0.1, 0.2, 0.3, 0.4, 0.5, 0.6, 0.7, 0.8, 0.9, 0.92,
      0.94, 0.96, 0.961, 0.9612, 0.9614, 0.9616 \}.
\end{array}
\end{displaymath}
Here we only take {those} resonances with
$m^2 \leq V_{{\rm{L}},{\rm{R}}}^{\text{max}}$ into account.

Because {the spectra of resonances or bound states for left- and right-handed fermions} are the same,
here and after we only discuss the spectrum of left-handed fermions.

\begin{figure}
\begin{center}
\includegraphics[width=0.45\textwidth]{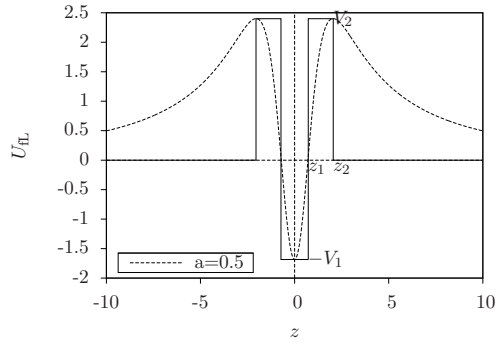}
\end{center}
\caption{ \label{fig:UfermionLeftflat05}
The fermion potential $U_{\rm{fL}}$ with $a=0$ and its volcano box potential approximation. The other parameters are set to $b=2$, and $c=1$}
\end{figure}

\begin{figure}
\begin{center}
\includegraphics[width=0.45\textwidth]{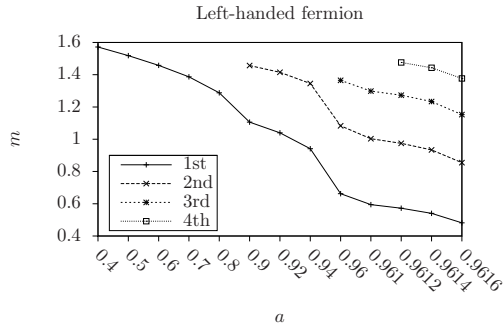}
\end{center}
\caption{ \label{fig:Resonances_m_a}
The variation of {the resonance} mass $m$
 with $a$ for left-handed fermions for flat brane. The other parameters are set to $b=2$, $c=1$, and $\eta=2$.}
\end{figure}

\begin{figure}
\begin{center}
\includegraphics[width=0.45\textwidth]{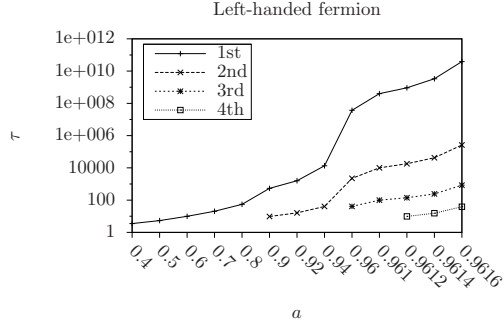}
\end{center}
\caption{ \label{fig:Resonances_t_a}
The variation of the resonance lifetime $\tau$ with $a$ for left-handed fermions for flat brane.  The other parameters are set to $b=2$, $c=1$, and $\eta=2$.}
\end{figure}

{
In order to see the effect of temperature on the fermion potentials $U_{\rm{fL}}$ and $U_{\rm{fL}}$
more intuitively,
we can take the volcano box potential approximation \cite{Csaki2000581} for the potentials. For example in Fig.
\ref{fig:UfermionLeftflat05}, the well's depth, height, and width are $V_1$, $V_2$, and $2 z_1$,
respectively. The width of  barrier is $|z_2-z_1|$. Now back to see the potential  $U_{\rm{fL}}$ in Fig. \ref{fig:cf_flat_Ufermion},
we can find that  with the increase of $a$ the width of barrier and the width of the
well are increasing, but the depth and height of the well are decreasing. Because the
variation of height of the well is small and all the KK modes with $m^2\geq0$, from the knowledge of quantum mechanics,
we can expect that with the increase of $a$ the number of the resonances is
increasing and their masses are decreasing. The numerical results support this  expectation.}
The variation of the resonance mass $m$ with $a$ is shown in Fig.~{\ref{fig:Resonances_m_a}}.
We can see the resonance mass decreases with the increase of
$a$. {While the resonance lifetime $\tau$ increases with $a$, which is plotted
in Fig.~{\ref{fig:Resonances_t_a}}.}

\begin{figure}
\begin{center}
\includegraphics[width=0.45\textwidth]{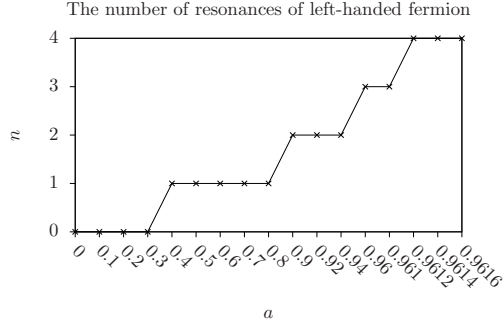}
\end{center}
\caption{ \label{fig:numResonances}
The variation of the number of resonances $n$ with $a$ for left-handed fermions for flat brane. The other parameters are set to $b=2$, $c=1$, and $\eta=2$.}
\end{figure}

The variation of the number of resonances $n$
with the temperature parameter $a$ is shown in Fig.~{\ref{fig:numResonances}}.
It can be seen that  the number of resonances $n$ is increasing
with the temperature parameter $a$, and there is a threshold
value for $a$ to produce more resonances. For example, for $a<0.4$ there is no resonance, and for $0.4\leq a<0.9$ there is a resonance. This is very similar to the phenomenon of the particle generation in colliders, namely, if we want to find more particles we need to increase
the center-mass energy and the generation of particles need the center-mass energy
greater than a threshold value.

\subsection{AdS brane}{\label{sec:ads}}

When $z\rightarrow 0$, the behaviors of the potentials of left- and right-handed fermions {for AdS brane}
{are similar to} the ones in the flat braneworld scenario, so we will not repeat them here.

But when $z\rightarrow\pm z_b$, the behaviors of $A$ {are} very different {from} the ones
in the cases of flat and dS branes, so we need to analyze the behaviors of {the potentials of left- and right-handed fermions} carefully.
When $z\rightarrow\pm  z_b$, the potentials turn into
\begin{subequations}
\begin{eqnarray}
  U_{\rm{fL}}&=& \eta e^{A}\phi_{\rm I}(\eta e^{A}\phi_{\rm I} - A'), \\
  U_{\rm{fR}}&=& \eta e^{A}\phi_{\rm I}(\eta e^{A}\phi_{\rm I} + A').
\end{eqnarray}
\end{subequations}
Inserting Eq. (\ref{A_adsinf}) into {the} above equations, we get
\begin{subequations}
\begin{eqnarray}
  U_{\rm{fL}}&=&
      \frac{-\Lambda_4  }{3} \frac{\eta}{\eta_0}
     \frac{\left( {\eta}/{\eta_0} -\cos(z')  \right)}{\sin^2(z')}, \\
  U_{\rm{fR}}&=&  \frac{-\Lambda_4  }{3} \frac{\eta}{\eta_0}
     \frac{\left( {\eta}/{\eta_0} +\cos(z')  \right)}{\sin^2(z')},
\end{eqnarray}
\end{subequations}
where {$z'={\sqrt{-\Lambda_4/3}(z_b-|z|)}$, $\eta_0\equiv \frac{\sqrt{2v_0}}{3\phi_0}$ with $v_{0}\equiv V(\phi_{\rm I}=\frac{\sqrt{3}}{2}\phi_0,\phi_{\rm R}=-\frac{1}{2}\phi_0)$}.
Then we can see that, if $\eta>\eta_0$,
\begin{subequations}
\begin{eqnarray}
 && U_{\rm{fL}}(z\rightarrow\pm  z_b)\rightarrow +\infty, \\
  &&U_{\rm{fR}}(z\rightarrow\pm  z_b)\rightarrow +\infty;
\end{eqnarray}
\end{subequations}
if $\eta=\eta_0$,
\begin{subequations}
\begin{eqnarray}
 && U_{\rm{fL}}(z\rightarrow\pm  z_b)\rightarrow-\frac{\Lambda _4}{6}, \\
  &&U_{\rm{fR}}(z\rightarrow\pm  z_b)\rightarrow +\infty;
\end{eqnarray}
\end{subequations}
if $0<\eta<\eta_0$,
\begin{subequations}
\begin{eqnarray}
 && U_{\rm{fL}}(z\rightarrow\pm  z_b)\rightarrow -\infty, \\
 && U_{\rm{fR}}(z\rightarrow\pm  z_b)\rightarrow +\infty.
\end{eqnarray}
\end{subequations}

\begin{figure*}
\begin{center}
\includegraphics[width=0.45\textwidth]{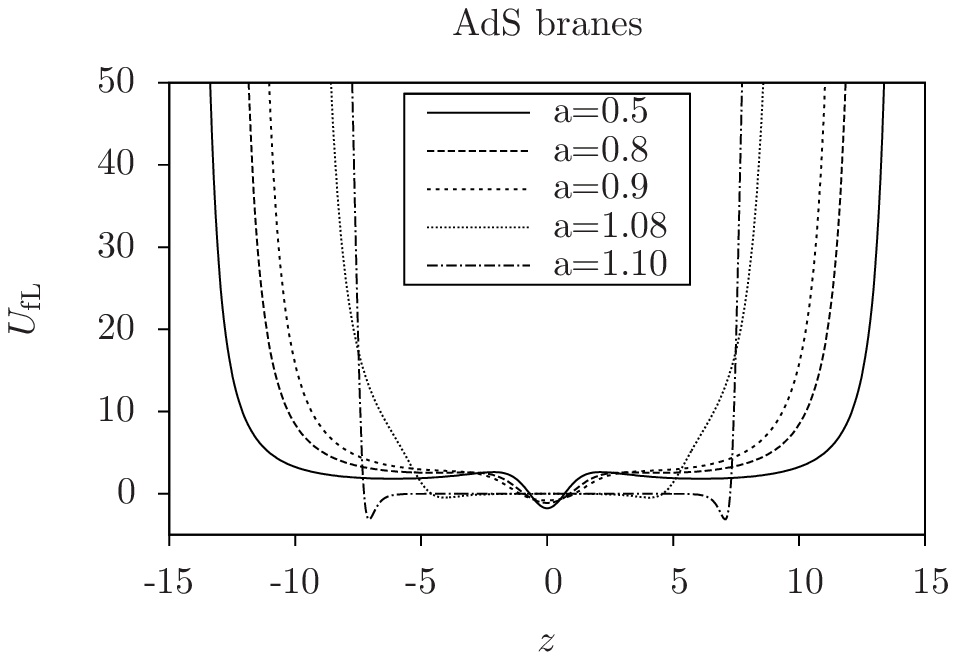}
\includegraphics[width=0.45\textwidth]{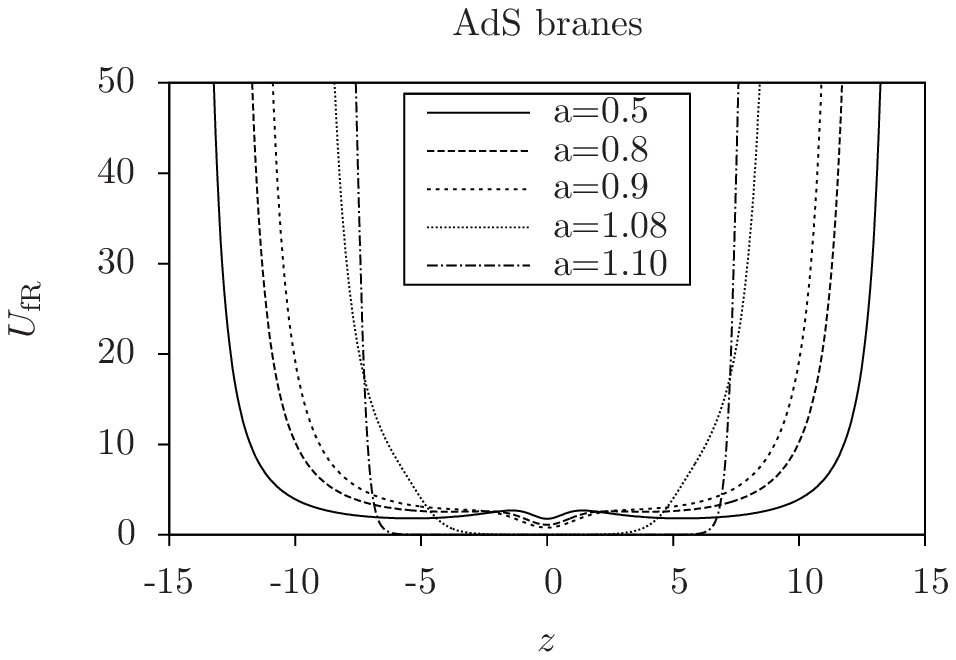}
\end{center}
\caption{\label{Uf_ads}
The plots of $U_{{\rm{fL}}}$ and $U_{{\rm{fR}}}$ {for AdS brane}.
The other parameters are set to $\eta=2$, $b=2$, $c=1$, and $\Lambda_4=-0.1$.}
\end{figure*}

In {the} cases of $\eta=\eta_0$ and $0<\eta<\eta_0$,
the behaviors of left- and right-hand fermions potentials
at $z\rightarrow \pm z_b$ are so different,
and we will not expect {the left- and right-handed fermions have the same KK modes spectra}.
{So} we only discuss the case of $\eta>\eta_0$.
The plots of the potentials of left- and right-hand fermions are shown {in Fig}.~{\ref{Uf_ads}}.

{From  Fig}.~{\ref{Uf_ads}} and {the above} analysis, we know
that, for $\eta>\eta_0$, all massive fermion KK modes are bound states
and there is a discrete mass spectrum for them.

We use the method proposed in Ref. \cite{FackBerghe1987} to calculate the fermion mass spectrum.
And we calculate the fermion mass spectra with {a set of values of} $a$:
\[a=\{0.2, 0.3, 0.4, 0.5, 0.6, 0.7, 0.8, 0.9, 1, 1.08, 1.10\},\]
 and the other parameters are set to $\eta=2$, $b=2$, $c=1$, and $\Lambda_4=-0.1$.
{The masses of 39 KK modes are obtained}.

\begin{figure}
\begin{center}
\includegraphics[width=0.5\textwidth]{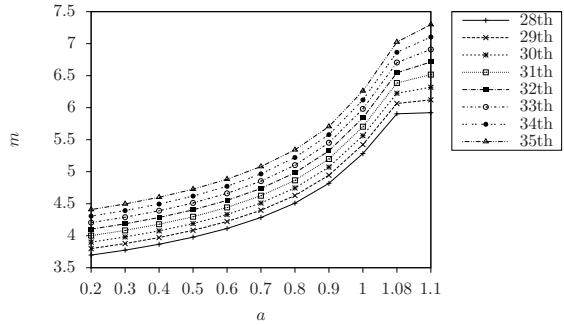}
\end{center}
\caption{ \label{LeftAdSMass3}
The $m-a$ curves of 28th-35th excited states for left-handed fermion with parameters
 $\eta=2$, $b=2$, $c=1$, and $\Lambda_4=-0.1$.}
\end{figure}

\begin{figure}
\begin{center}
\includegraphics[width=0.5\textwidth]{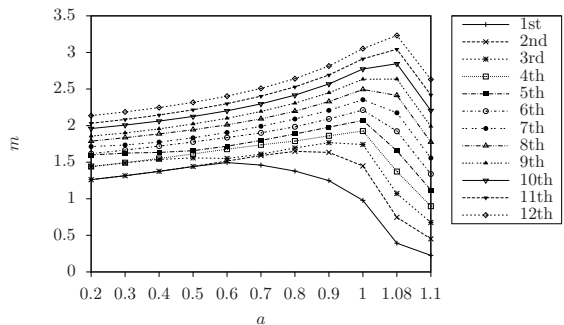}
\end{center}
\caption{ \label{LeftAdSMass2}
The $m-a$ curves of 1st-12th excited states for left-handed fermions {for AdS brane} with parameters
 $\eta=2$, $b=2$, $c=1$, and $\Lambda_4=-0.1$.}
\end{figure}

{
From Figure \ref{Uf_ads}, we can see that the profile of potentials $U_{{\rm{fL}}}$ and $U_{{\rm{fR}}}$
is similar to that of the infinite square-well potential.
And we know that a particle in the infinite square-well potential will hold the the
energy of $E_n\propto \frac{n^2}{d^2}$ [$n=1, 2, 3, \ldots$], where $d$ is the width of well,
and $E_n$ decreases with $d$.
Figure \ref{Uf_ads} shows that with the increase of $a$ the width of the potentials is decreasing,
so we except that the masses of the KK modes increase with $a$.
But this is not all consistent to our numerical results.
For the 28th excited state and the ones above it, the {masses} of KK modes
increase with $a$. For example, we show the $m-a$ curves of 28th-35th excited states in
Fig.~{\ref{LeftAdSMass3}}. But the ones of below the 28th excited state
masses of the KK modes do not vary monotonically with the increase of $a$ (see Fig. \ref{LeftAdSMass2}).
 This is because with the increase of
$a$ that the values of the bottom of potentials are also not varying  monotonically with the increase of $a$.
For example the first excited state (left-handed fermion) mass $m_1(a=0.2)$ greater than $m_1(a=1.1)$
 is because the value of the bottom of $U_{{\rm{fL}}}$ at $a=0.2$ is greater than the one
 of $U_{{\rm{fL}}}$ at $a=1.1$.
}


\section{Conclusion and discussion}\label{sec:conclusion}

With the model used in this paper, we {found}
the effects of temperature on flat and dS branes are {similar},
except {the existence of the critical temperature}.
In the case of flat brane, there is a critical temperature. {While it does not exist for dS brane.}
The thickness of flat brane is uncertain at the critical temperature.
{We think this is related to the potential given in (\ref{V(Phi)}). If we choose the
$\phi^4$ model of a real scalar field}, there will not be such a critical temperature even for flat brane.

In the case of AdS brane, we found that
{the critical temperature} is greater than the one in 5D flat spacetime.
This {is just opposite to} the case of flat brane, for which
the critical temperature is less than the one
in 5D flat spacetime \cite{Campos200288}.

The variation of the temperature parameter will affect the spectrum of
fermion KK modes. First, in the cases of flat and dS branes, the spectrum is continuous.
We found that, for some parameters, there are fermion resonances on the brane.
The fermion resonances can be thought as quasi-localized fermions with finite lifetime staying on the brane.
The number of fermion resonances is finite and increases with the temperature parameter $a$,
{so does} the resonance lifetime.
{This} means that the brane with a bigger value of temperature parameter can trap more {quasi-localized fermions on it}.
The masses of the quasi-localized fermions decrease with the increase of $a$.
Second, in the case of AdS brane, if the Yukawa coupling is larger than some critical coupling constant, then there is a discrete spectrum of fermion KK modes, and all the KK modes (four-dimensional fermions) are localized on the brane. But the masses of fermion KK modes do not vary monotonically
with the increase of the temperature parameter.

From the model discussed in this paper we can see: (a) In the case of AdS brane, the fermions will be
bounded on our brane permanently and the number of fermions is infinite. But the model with an AdS brane is unreal.
This is because we know that in our real universe the 4D cosmological constant is positive.
(b) Corresponding to the AdS brane, the flat and dS branes are more similar to our real universe.
In these cases the fermions are quasi-localized states, except the zero mode,
with a finite lifetime staying on the branes. This means that the fermions can dissipate into the fifth dimension
just like the quasi-localized gauge field \cite{Dvali2001} and the quasi-localized gravitons
 \cite{CsakiErlichHollowood2000,DvaliGabadadzePorrati2000,GregoryRubakovSibiryakov2000a,GregoryRubakovSibiryakov2000}.
From the point of view of 4D observer,
one will have the chance to find energy non-conservation in the collider.

{
If we assume that the temperature parameter $a$ is a monotone increasing function of temperature $T$,
then, for a collider experiment, the number of the produced new fermion resonances
will increase with the center-of-mass energy of beams.
This is consistent to results of the experiment.
Further, our another result that the lifetime of fermions increase with the temperature, would be examined if the collider experiment find the signal of fermions dissipating into the fifth dimension.
}

\section*{Acknowledgments}

This work was supported by the Program for New Century Excellent
Talents in University, the Huo Ying-Dong Education Foundation
of Chinese Ministry of Education (No. 121106), the National Natural Science Foundation of
China (No. 11075065 and No. 11005054), the Doctoral Program Foundation of
Institutions of Higher Education of China (No. 20090211110028),
the Key Project of the Chinese Ministry of Education (No. 109153), and
the Natural Science Foundation of Gansu Province, China (No. 096RJZA055). Y.Q. Wang was supported by
the Fundamental Research Fund
for Physics and Mathematics of Lanzhou University (No. LZULL200912).
Z.H. Zhao was supported by the Scholarship Award for Excellent Doctoral
Student granted by Ministry of Education.

\section{Appendix}\label{appendixe}

Here we give the discussion about the relation between $a_*$ and $a_c$.

First at the point of $z=0$, substituting $A(0)=A'(0)=\phi_{\rm I}(0)=\phi'_{\rm R}(0)=0$ into Eq. (\ref{C_5_cf}) we get:
 \begin{equation}
   2\Lambda_4+\frac{1}{2}\phi'_{\rm I}(0)-V_0=0,\label{262}
 \end{equation}
where
\begin{eqnarray}
  V_0&=&V(\phi_{\rm I}(0),\phi_{\rm R}(0))\nonumber\\
  &=&a\phi^2_{\rm R}(0)-b\phi^3_{\rm R}(0)+c\phi^4_{\rm R}(0)+C_5.\label{V0}
  \end{eqnarray}
Because $\phi_{\rm R}(z)$ is an even function of $z$, we can Taylor expand $\phi_{\rm R}(z)$ around the point of $z=0$:
 \begin{equation}
   \phi_{\rm R}(z)=d_0+d_2 z^2+d_4z^4+\cdots.
 \end{equation}
So $\phi_{\rm R}(0)=d_0$. Substituting this into Eq. (\ref{V0}) we get
\begin{eqnarray}
  V_0=a d_0^2-b d_0^3+c d_0^4+C_5.\label{V02}
\end{eqnarray}
And Eq. (\ref{262}) is rewritten as
 \begin{equation}
   2\Lambda_4+\frac{1}{2}\phi'_{\rm I}(0)=a d_0^2-b d_0^3+c d_0^4+C_5. \label{2622}
 \end{equation}

 Further, when $z$ tends to its right boundary ( for flat and dS branes it is $\infty$ but for AdS brane it is $z_b$,
 here we use symbol $\infty$ to stand for it).
 \begin{eqnarray}
  V_{\infty}=a |\Phi(\infty)|^2-b \phi_{\rm R}(\infty)(\phi_{\rm R}(\infty)^2-3\phi_{\rm I}(\infty)^2)+c |\Phi(\infty)|^4 +C_5.\label{Vinfty}
\end{eqnarray}
Substituting Eq. (\ref{28}) into the above equation leads to
 \begin{eqnarray}
  V_{\infty}=a \phi_{0}^2-b \phi_{0}^3+c \phi_{0}^4 +C_5, \label{Vinfty2}
\end{eqnarray}
where $\phi_0=\frac{3b}{8c}\big(1+\sqrt{(9-8a)/a_c}/{3}\big)$. So
 \begin{eqnarray}
  C_5=V_{\infty}-(a \phi_{0}^2-b \phi_{0}^3+c \phi_{0}^4).\label{C5}
\end{eqnarray}
Substituting the above expression for $C_5$ into Eq. (\ref{2622}) gives
 \begin{eqnarray}
  a=(b\phi_{0}-c\phi_{0}^2)\frac{\phi_{0}^2}{\phi_{0}^2-d_0^2}-\frac{b-c d_0}{\phi_{0}^2-d_0^2}d_0^3
  -\frac{2\Lambda_4}{\phi_{0}^2-d_0^2}+\frac{V_{\infty}}{\phi_{0}^2-d_0^2}
  -\frac{1}{2}\frac{\phi'_{\rm I}(0)}{\phi_{0}^2-d_0^2}.\label{a}
\end{eqnarray}
When $\phi'_{\rm I}(0)\rightarrow 0$, $a\rightarrow a_*$.
 We assume that $d^2_0 \ll \phi^2_0$
(this assumption is supported by the numerical results
which can be seen from Figs. \ref{fig_cf_flat_Phi}, \ref{fig_cf_flat_Phi_critical},
and  \ref{fig:phiIupCritical_ads}). Neglecting the terms containing square and higher powers of $d_0$,
Eq. (\ref{a}) reads
 \begin{eqnarray}
  a_*=(b\phi_{0}-c\phi_{0}^2)
  -\frac{1}{\phi_{0}^2}(2\Lambda_4-V_{\infty})
  .\label{a2}
\end{eqnarray}
Further we take $a_*=a_c(1-\delta)$ [$a_c=b^2/(4c)=1$] and assume that $|\delta|\ll 1$. So
$\phi_0=\frac{3b}{8c}\big(1+\frac{\sqrt{1+8\delta}}{3}\big)$. The Taylor expansions of
Eq. (\ref{a2}) respect to $\delta$ is
 \begin{eqnarray}
a_*=a_c(1-\delta^2+4\delta^3)-\frac{4 c^2}{b^2}(2\Lambda_4
-V_{\infty})(1-2\delta+7\delta^2-32\delta^3)+ \mathcal{O}(\delta )^4.\label{a3}
\end{eqnarray}
After neglecting the terms containing square and higher powers of $\delta$, Eq. (\ref{a3}) reads
 \begin{eqnarray}
a_*=a_c-\frac{4 c^2}{b^2}(2\Lambda_4
-V_{\infty})(1-2\delta).\label{a4}
\end{eqnarray}

So in the case of flat brane, we have $\Lambda_4=0$ and $V_{\infty}\leq0$ (this can be found from Eq. (\ref{Aflat})),
 which leads to $a_*\leq a_c$.

{
But in the cases of dS and AdS branes we cannot confirm whether $a_*>a_c$ or $a_*\leq a_c$.
That is because that, in the  case of dS brane $\Lambda_4>0$ and $V_{\infty}>0$ (this can be found from Eq. (\ref{A_dS})),
and in the case of AdS brane $\Lambda_4<0$ and $V_{\infty}<0$ (this can be found from Eq. (\ref{A_adsinf})), we are not able to determine whether $(2\Lambda_4-V_{\infty})>0$ or $\leq 0$. Even in the case of dS brane,  we cannot confirm the existence of the $a_*$ using the numerical method.}

\begin{figure}[h]
\begin{center}
\includegraphics[width=0.5\textwidth]{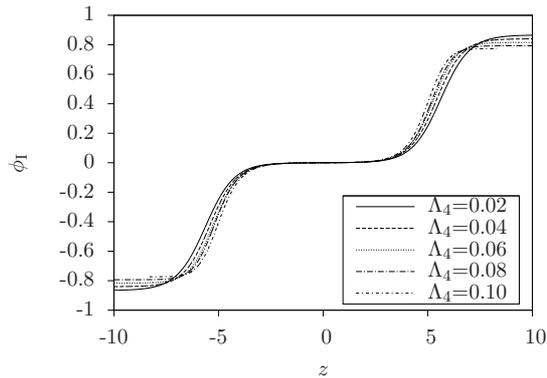}
\end{center}
\caption{ \label{phiICritical_ads}
The solutions of $\phi_{\mathrm I}$ with different values of $\Lambda_4$ (given in Table 1)
corresponding to $a=a_*$ for the case of AdS brane.
The other parameters are set to $b=2$ and  $c=1$..}
\end{figure}

\begin{figure}[h]
\begin{center}
\includegraphics[width=0.5\textwidth]{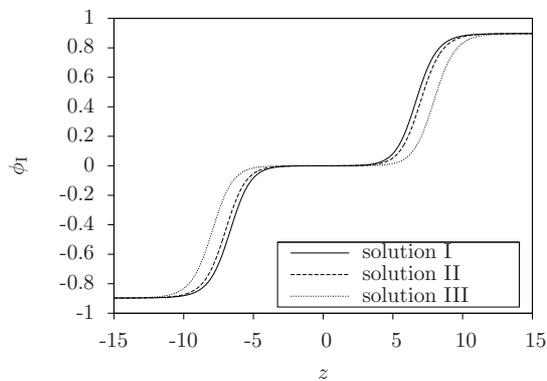}
\end{center}
\caption{ \label{phiIs_critical_flat}
Three solutions of $\phi_{\mathrm I}$ with the same value of
$a_*=0.9617$ for the case of flat brane.
The other parameters are set to $b=2$ and  $c=1$.}
\end{figure}

{
In the above discussion we have made use of two assumptions: $|\delta|\ll 1$ and $d_0^2 \ll \phi_0^2$.
This will weak the efficiency of our conclusion.
So here we make use of the numerical method to check the consistency between the two assumptions and  Eq. (\ref{a4}).
From Eq. (\ref{a4}) we have
\begin{eqnarray}
\delta=1-\frac{b^2}{b^2+4c^2(2\Lambda_4-V_{\infty})}. \label{delta}
\end{eqnarray}
Using the numerical method, we have obtained a set of values of $V_{\infty}$, $d_0$ and $\delta$  corresponding to $a=a_*$.
All results are shown in Table \ref{tab1} for the case of AdS brane (the corresponding solutions of $\phi_{\rm I}$ are shown in Fig. \ref{phiICritical_ads})
and Table \ref{tab2} for the case of flat brane (the corresponding solutions of $\phi_{\rm I}$ are shown in Fig. \ref{phiIs_critical_flat}), which show that the value of $a_c-a_*$ is almost the same with the $\delta$ obtained from Eq. (\ref{a4}).
So the consistency is satisfied.
}

\begin{table}[h]
\caption{The numerical results of $V_{\infty}$, $d_0$ and $\delta$  corresponding to $a=a_*$ for the case of AdS brane with $b=2$, $c=1$, and $a_c=1$.\label{tab1}}
\begin{center}
\begin{tabular}{cccccc}
  \hline\hline
   $V_{\infty}$&$d_0$& $a_*$ & $a_c-a_*$ &$\delta$ \\
   \hline
  -0.041001 &-0.000327 & 0.999 &  0.001 & 0.000999 \\
  -0.052821 &-0.000350 & 1.028 & -0.028 & -0.0279383\\
  -0.071759 &-0.000366 & 1.051 & -0.051 & -0.0506862\\
  -0.095345 &-0.000339 & 1.070 & -0.070 & -0.0691242\\
  -0.123884 &-0.000382 & 1.084 & -0.084 & -0.082387 \\
  \hline\hline
\end{tabular}
\end{center}
\end{table}

\begin{table}[h]
\caption{The numerical results of $V_{\infty}$, $d_0$ and $\delta$ corresponding to $a=a_*$ for the case of flat brane with $b=2$, $c=1$, and $a_c=1$. The solutions I, II, and III of $\phi_{\mathrm I}$ are showed in Fig. (\ref{phiIs_critical_flat}), and they correspond to
the same value of $a_*=0.9617$.
  \label{tab2}}
\begin{center}
\begin{tabular}{cccccc}
  \hline\hline
  $\phi_{\mathrm I}$ &$V_{\infty}$&$d_0$& $a_*$ & $a_c-a_*$ &$\delta$ \\
   \hline
  solution I  & -0.039713 &-0.000103 & 0.9617 & 0.0383 & 0.038196 \\
  solution II & -0.039714 &-0.000064 & 0.9617 & 0.0383 & 0.038197\\
  solution III& -0.039714 &-0.000017 & 0.9617 & 0.0383 & 0.038197\\
  \hline\hline
\end{tabular}
\end{center}
\end{table}

\bibliographystyle{JHEP}
\bibliography{bibs}

\providecommand{\href}[2]{#2}\begingroup\raggedright\begin{thebibliography}{10}

\bibitem{Arkani-Hamed1998429}
N.~Arkani-Hamed, S.~Dimopoulos, and G.~R. Dvali, {\it The hierarchy problem and
  new dimensions at a millimeter},  {\em Phys. Lett. B} {\bf 429} (1998) 263,
  [\href{http://xxx.lanl.gov/abs/hep-ph/9803315}{{\tt hep-ph/9803315}}].

\bibitem{AntoniadisDimopoulosDvali1998}
I.~Antoniadis, S.~Dimopoulos, and G.~R. Dvali, {\it Millimeter range forces in
  superstring theories with weak- scale compactification},  {\em Nucl. Phys. B}
  {\bf 516} (1998) 70, [\href{http://xxx.lanl.gov/abs/hep-ph/9710204}{{\tt
  hep-ph/9710204}}].

\bibitem{AntoniadisArkani-HamedDimopoulosDvali1998}
I.~Antoniadis, N.~Arkani-Hamed, S.~Dimopoulos, and G.~R. Dvali, {\it New
  dimensions at a millimeter to a fermi and superstrings at a tev},  {\em Phys.
  Lett. B} {\bf 436} (1998) 257,
  [\href{http://xxx.lanl.gov/abs/hep-ph/9804398}{{\tt hep-ph/9804398}}].

\bibitem{Randall199983}
L.~Randall and R.~Sundrum, {\it A large mass hierarchy from a small extra
  dimension},  {\em Phys. Rev. Lett.} {\bf 83} (1999) 3370,
  [\href{http://xxx.lanl.gov/abs/hep-ph/9905221}{{\tt hep-ph/9905221}}].

\bibitem{Randall199983a}
L.~Randall and R.~Sundrum, {\it An alternative to compactification},  {\em
  Phys. Rev. Lett.} {\bf 83} (1999) 4690,
  [\href{http://xxx.lanl.gov/abs/hep-th/9906064}{{\tt hep-th/9906064}}].

\bibitem{Bajc2000}
B.~Bajc and G.~Gabadadze, {\it Localization of matter and cosmological constant
  on a brane in anti de sitter space},  {\em Phys. Lett. B} {\bf 474} (2000)
  282, [\href{http://xxx.lanl.gov/abs/hep-th/9912232}{{\tt hep-th/9912232}}].

\bibitem{Randjbar-Daemi2000}
S.~Randjbar-Daemi and M.~E. Shaposhnikov, {\it Fermion zero-modes on
  brane-worlds},  {\em Phys. Lett. B} {\bf 492} (2000) 361,
  [\href{http://xxx.lanl.gov/abs/hep-th/0008079}{{\tt hep-th/0008079}}].

\bibitem{Pomarol2000}
A.~Pomarol, {\it Gauge bosons in a five-dimensional theory with localized
  gravity},  {\em Phys. Lett. B} {\bf 486} (2000) 153,
  [\href{http://xxx.lanl.gov/abs/hep-ph/9911294}{{\tt hep-ph/9911294}}].

\bibitem{CasadioGruppusoVenturi2000}
R.~Casadio, A.~Gruppuso, and G.~Venturi, {\it Electromagnetic contributions to
  lepton g-2 in a thick brane-world},  {\em Phys. Lett. B} {\bf 495} (2000)
  378, [\href{http://xxx.lanl.gov/abs/hep-th/0010065}{{\tt hep-th/0010065}}].

\bibitem{CasadioGruppuso2001}
R.~Casadio and A.~Gruppuso, {\it Discrete symmetries and localization in a
  brane-world},  {\em Phys. Rev. D} {\bf 64} (2001) 025020,
  [\href{http://xxx.lanl.gov/abs/hep-th/0103200}{{\tt hep-th/0103200}}].

\bibitem{Akhmedov2001}
E.~K. Akhmedov, {\it Dynamical localization of gauge fields on a brane},  {\em
  Phys. Lett. B} {\bf 521} (2001) 79,
  [\href{http://xxx.lanl.gov/abs/hep-th/0107223}{{\tt hep-th/0107223}}].

\bibitem{Dvali2001}
G.~R. Dvali, G.~Gabadadze, and M.~A. Shifman, {\it (quasi)localized gauge field
  on a brane: Dissipating cosmic radiation to extra dimensions?},  {\em Phys.
  Lett. B} {\bf 497} (2001) 271,
  [\href{http://xxx.lanl.gov/abs/hep-th/0010071}{{\tt hep-th/0010071}}].

\bibitem{Ghoroku2002}
K.~Ghoroku and A.~Nakamura, {\it Massive vector trapping as a gauge boson on a
  brane},  {\em Phys. Rev. D} {\bf 65} (2002) 084017,
  [\href{http://xxx.lanl.gov/abs/hep-th/0106145}{{\tt hep-th/0106145}}].

\bibitem{Ringeval200265}
C.~Ringeval, P.~Peter, and J.-P. Uzan, {\it Localization of massive fermions on
  the brane},  {\em Phys. Rev. D} {\bf 65} (2002) 044016,
  [\href{http://xxx.lanl.gov/abs/hep-th/0109194}{{\tt hep-th/0109194}}].

\bibitem{Koley200522}
R.~Koley and S.~Kar, {\it Scalar kinks and fermion localization in warped
  spacetimes},  {\em Class. Quant. Grav.} {\bf 22} (2005) 753,
  [\href{http://xxx.lanl.gov/abs/hep-th/0407158}{{\tt hep-th/0407158}}].

\bibitem{Oda2003}
I.~Oda, {\it Gravitational localization of all local fields on the brane},
  {\em Phys. Lett. B} {\bf 571} (2003) 235,
  [\href{http://xxx.lanl.gov/abs/hep-th/0307119}{{\tt hep-th/0307119}}].

\bibitem{Melfo2006}
A.~Melfo, N.~Pantoja, and J.~D. Tempo, {\it Fermion localization on thick
  branes},  {\em Phys. Rev. D} {\bf 73} (2006) 044033,
  [\href{http://xxx.lanl.gov/abs/hep-th/0601161}{{\tt hep-th/0601161}}].

\bibitem{Liu2007a}
Y.-X. Liu, L.~Zhao, X.-H. Zhang, and Y.-S. Duan, {\it Fermions in self-dual
  vortex background on a string-like defect},  {\em Nucl. Phys. B} {\bf 785}
  (2007) 234, [\href{http://xxx.lanl.gov/abs/0704.2812}{{\tt
  arXiv:0704.2812}}].

\bibitem{Liu2007}
Y.-X. Liu, L.~Zhao, and Y.-S. Duan, {\it Localization of fermions on a
  string-like defect},  {\em J. High Energy Phys.} {\bf 04} (2007) 097,
  [\href{http://xxx.lanl.gov/abs/hep-th/0701010}{{\tt hep-th/0701010}}].

\bibitem{Davies2007}
R.~Davies and D.~P. George, {\it Fermions, scalars and randall-sundrum gravity
  on domain- wall branes},  {\em Phys. Rev. D} {\bf 76} (2007) 104010,
  [\href{http://xxx.lanl.gov/abs/0705.1391}{{\tt arXiv:0705.1391}}].

\bibitem{Slatyer2007}
T.~R. Slatyer and R.~R. Volkas, {\it Cosmology and fermion confinement in a
  scalar-field- generated domain wall brane in five dimensions},  {\em J. High
  Energy Phys.} {\bf 04} (2007) 062,
  [\href{http://xxx.lanl.gov/abs/hep-ph/0609003}{{\tt hep-ph/0609003}}].

\bibitem{KoleyMitraSenGupta2010}
R.~Koley, J.~Mitra, and S.~SenGupta, {\it Scalar kaluza-klein modes in a
  multiply warped braneworld},  {\em Europhys. Lett.} {\bf 91} (2010) 31001,
  [\href{http://xxx.lanl.gov/abs/1001.2666}{{\tt arXiv:1001.2666}}].

\bibitem{Liu200878}
Y.-X. Liu, L.-D. Zhang, L.-J. Zhang, and Y.-S. Duan, {\it Fermions on thick
  branes in background of sine-gordon kinks},  {\em Phys. Rev. D} {\bf 78}
  (2008) 065025, [\href{http://xxx.lanl.gov/abs/0804.4553}{{\tt
  arXiv:0804.4553}}].

\bibitem{Liu200808}
Y.-X. Liu, L.-D. Zhang, S.-W. Wei, and Y.-S. Duan, {\it Localization and mass
  spectrum of matters on weyl thick branes},  {\em J. High Energy Phys.} {\bf
  08} (2008) 041, [\href{http://xxx.lanl.gov/abs/0803.0098}{{\tt
  arXiv:0803.0098}}].

\bibitem{Liu200802}
Y.-X. Liu, X.-H. Zhang, L.-D. Zhang, and Y.-S. Duan, {\it Localization of
  matters on pure geometrical thick branes},  {\em J. High Energy Phys.} {\bf
  02} (2008) 067, [\href{http://xxx.lanl.gov/abs/0708.0065}{{\tt
  arXiv:0708.0065}}].

\bibitem{Guerrero2009}
R.~Guerrero, A.~Melfo, N.~Pantoja, and R.~O. Rodriguez, {\it Gauge field
  localization on brane worlds},  {\em Phys. Rev. D} {\bf 81} (2010) 086004,
  [\href{http://xxx.lanl.gov/abs/0912.0463}{{\tt arXiv:0912.0463}}].

\bibitem{Zhao2009}
Z.-H. Zhao, Y.-X. Liu, and H.-T. Li, {\it Fermion localization on asymmetric
  two-field thick branes},  {\em Class. Quant. Grav.} {\bf 27} (2010) 185001,
  [\href{http://xxx.lanl.gov/abs/0911.2572}{{\tt arXiv:0911.2572}}].

\bibitem{Liu2009a}
Y.-X. Liu, C.-E. Fu, L.~Zhao, and Y.-S. Duan, {\it Localization and mass
  spectra of fermions on symmetric and asymmetric thick branes},  {\em Phys.
  Rev. D} {\bf 80} (2009) 065020,
  [\href{http://xxx.lanl.gov/abs/0907.0910}{{\tt arXiv:0907.0910}}].

\bibitem{Liu2010}
Y.-X. Liu, H.~Guo, C.-E. Fu, and J.-R. Ren, {\it Localization of matters on
  anti-de sitter thick branes},  {\em J. High Energy Phys.} {\bf 02} (2010)
  080, [\href{http://xxx.lanl.gov/abs/0907.4424}{{\tt arXiv:0907.4424}}].

\bibitem{CorreaDutraHott2010}
R.~A.~C. Correa, A.~d.~S. Dutra, and M.~B. Hott, ``Fermion localization on
  degenerate and critical branes.'' 2010.

\bibitem{Castro2011}
L.~B. Castro, {\it Fermion localization on two-field thick branes},  {\em Phys.
  Rev. D} {\bf 83} (2011) 045002,
  [\href{http://xxx.lanl.gov/abs/1008.3665}{{\tt arXiv:1008.3665}}].

\bibitem{CastroMeza2010}
L.~B. Castro and L.~A. Meza, ``Fermion localization on branes with generalized
  dynamics.'' 2010.

\bibitem{FrereLibanovTroitsky2001}
J.~M. Fr\`{e}re, M.~V. Libanov, and S.~V. Troitsky, {\it Three generations on a
  local vortex in extra dimensions},  {\em Phys. Lett. B} {\bf 512} (2001) 169,
  [\href{http://xxx.lanl.gov/abs/hep-ph/0012306}{{\tt hep-ph/0012306}}].

\bibitem{LibanovTroitsky2001}
M.~V. Libanov and S.~V. Troitsky, {\it Three fermionic generations on a
  topological defect in extra dimensions},  {\em Nucl. Phys. B} {\bf 599}
  (2001) 319, [\href{http://xxx.lanl.gov/abs/hep-ph/0011095}{{\tt
  hep-ph/0011095}}].

\bibitem{Neronov2002}
A.~Neronov, {\it Fermion masses and quantum numbers from extra dimensions},
  {\em Phys. Rev. D} {\bf 65} (2002) 044004,
  [\href{http://xxx.lanl.gov/abs/gr-qc/0106092}{{\tt gr-qc/0106092}}].

\bibitem{AguilarSingleton2006}
S.~Aguilar and D.~Singleton, {\it Fermion generations, masses and mixings in a
  6d brane model},  {\em Phys. Rev. D} {\bf 73} (2006) 085007,
  [\href{http://xxx.lanl.gov/abs/hep-th/0602218}{{\tt hep-th/0602218}}].

\bibitem{GogberashviliMidodashviliSingleton2007}
M.~Gogberashvili, P.~Midodashvili, and D.~Singleton, {\it Fermion generations
  from 'apple-shaped' extra dimensions},  {\em JHEP} {\bf 08} (2007) 033,
  [\href{http://xxx.lanl.gov/abs/0706.0676}{{\tt arXiv:0706.0676}}].

\bibitem{GuoMa2008}
Z.-q. Guo and B.-Q. Ma, {\it Fermion families from two layer warped extra
  dimensions},  {\em J. High Energy Phys.} {\bf 08} (2008) 065,
  [\href{http://xxx.lanl.gov/abs/0808.2136}{{\tt arXiv:0808.2136}}].

\bibitem{GuoMa2009}
Z.-Q. Guo and B.-Q. Ma, {\it A model of fermion masses and mixings triggered by
  family problem in warped extra dimensions},  {\em J. High Energy Phys.} {\bf
  09} (2009) 091, [\href{http://xxx.lanl.gov/abs/0909.4355}{{\tt
  arXiv:0909.4355}}].

\bibitem{Brevik2001599}
I.~H. Brevik, K.~A. Milton, S.~Nojiri, and S.~D. Odintsov, {\it Quantum
  (in)stability of a brane-world ads$_5$ universe at nonzero temperature},
  {\em Nucl. Phys. B} {\bf 599} (2001) 305,
  [\href{http://xxx.lanl.gov/abs/hep-th/0010205}{{\tt hep-th/0010205}}].

\bibitem{Campos200288}
A.~Campos, {\it Critical phenomena of thick branes in warped spacetimes},  {\em
  Phys. Rev. Lett.} {\bf 88} (2002) 141602,
  [\href{http://xxx.lanl.gov/abs/hep-th/0111207}{{\tt hep-th/0111207}}].

\bibitem{Bazeia200411}
D.~Bazeia, F.~A. Brito, and A.~R. Gomes, {\it Locally localized gravity and
  geometric transitions},  {\em J. High Energy Phys.} {\bf 11} (2004) 070,
  [\href{http://xxx.lanl.gov/abs/hep-th/0411088}{{\tt hep-th/0411088}}].

\bibitem{Bonjour1999}
F.~Bonjour, C.~Charmousis, and R.~Gregory, {\it Thick domain wall universes},
  {\em Class. Quant. Grav.} {\bf 16} (1999) 2427,
  [\href{http://xxx.lanl.gov/abs/gr-qc/9902081}{{\tt gr-qc/9902081}}].

\bibitem{Goldberger1999}
W.~D. Goldberger and M.~B. Wise, {\it Modulus stabilization with bulk fields},
  {\em Phys. Rev. Lett.} {\bf 83} (1999) 4922,
  [\href{http://xxx.lanl.gov/abs/hep-ph/9907447}{{\tt hep-ph/9907447}}].

\bibitem{Csaki2000581}
C.~Cs\'{a}ki, J.~Erlich, T.~J. Hollowood, and Y.~Shirman, {\it Universal
  aspects of gravity localized on thick branes},  {\em Nucl. Phys. B} {\bf 581}
  (2000) 309, [\href{http://xxx.lanl.gov/abs/hep-th/0001033}{{\tt
  hep-th/0001033}}].

\bibitem{DeWolfe200062}
O.~DeWolfe, D.~Z. Freedman, S.~S. Gubser, and A.~Karch, {\it Modeling the fifth
  dimension with scalars and gravity},  {\em Phys. Rev. D} {\bf 62} (2000)
  046008, [\href{http://xxx.lanl.gov/abs/hep-th/9909134}{{\tt
  hep-th/9909134}}].

\bibitem{Gremm2000478}
M.~Gremm, {\it Four-dimensional gravity on a thick domain wall},  {\em Phys.
  Lett. B} {\bf 478} (2000) 434,
  [\href{http://xxx.lanl.gov/abs/hep-th/9912060}{{\tt hep-th/9912060}}].

\bibitem{Ichinose200118}
S.~Ichinose, {\it Some properties of domain wall solution in the randall-
  sundrum model},  {\em Class. Quant. Grav.} {\bf 18} (2001) 5239,
  [\href{http://xxx.lanl.gov/abs/hep-th/0107254}{{\tt hep-th/0107254}}].

\bibitem{Ichinose200118a}
S.~Ichinose, {\it A solution of the randall-sundrum model and the mass
  hierarchy problem},  {\em Class. Quant. Grav.} {\bf 18} (2001) 421.

\bibitem{Kehagias2001504}
A.~Kehagias and K.~Tamvakis, {\it Localized gravitons, gauge bosons and chiral
  fermions in smooth spaces generated by a bounce},  {\em Phys. Lett. B} {\bf
  504} (2001) 38, [\href{http://xxx.lanl.gov/abs/hep-th/0010112}{{\tt
  hep-th/0010112}}].

\bibitem{Gregory2002}
R.~Gregory and A.~Padilla, {\it Nested braneworlds and strong brane gravity},
  {\em Phys. Rev. D} {\bf 65} (2002) 084013,
  [\href{http://xxx.lanl.gov/abs/hep-th/0104262}{{\tt hep-th/0104262}}].

\bibitem{Gregory2002a}
R.~Gregory and A.~Padilla, {\it Braneworld instantons},  {\em Class. Quant.
  Grav.} {\bf 19} (2002) 279,
  [\href{http://xxx.lanl.gov/abs/hep-th/0107108}{{\tt hep-th/0107108}}].

\bibitem{Kobayashi2002}
S.~Kobayashi, K.~Koyama, and J.~Soda, {\it Thick brane worlds and their
  stability},  {\em Phys. Rev. D} {\bf 65} (2002) 064014,
  [\href{http://xxx.lanl.gov/abs/hep-th/0107025}{{\tt hep-th/0107025}}].

\bibitem{Bazeia200391}
D.~Bazeia, J.~Menezes, and R.~Menezes, {\it New global defect structures},
  {\em Phys. Rev. Lett.} {\bf 91} (2003) 241601.

\bibitem{Bronnikov2003}
K.~A. Bronnikov and B.~E. Meierovich, {\it A general thick brane supported by a
  scalar field},  {\em Grav. Cosmol.} {\bf 9} (2003) 313,
  [\href{http://xxx.lanl.gov/abs/gr-qc/0402030}{{\tt gr-qc/0402030}}].

\bibitem{Eto200368}
M.~Eto and N.~Sakai, {\it Solvable models of domain walls in $n=1$
  supergravity},  {\em Phys. Rev. D} {\bf 68} (2003) 125001.

\bibitem{Melfo2003}
A.~Melfo, N.~Pantoja, and A.~Skirzewski, {\it Thick domain wall spacetimes with
  and without reflection symmetry},  {\em Phys. Rev. D} {\bf 67} (2003) 105003,
  [\href{http://xxx.lanl.gov/abs/gr-qc/0211081}{{\tt gr-qc/0211081}}].

\bibitem{Bazeia2004}
D.~Bazeia, C.~Furtado, and A.~R. Gomes, {\it Brane structure from scalar field
  in warped spacetime},  {\em J. Cosmol. Astropart. Phys.} {\bf 0402} (2004)
  002, [\href{http://xxx.lanl.gov/abs/hep-th/0308034}{{\tt hep-th/0308034}}].

\bibitem{Bazeia200405}
D.~Bazeia and A.~R. Gomes, {\it Bloch brane},  {\em J. High Energy Phys.} {\bf
  05} (2004) 012, [\href{http://xxx.lanl.gov/abs/hep-th/0403141}{{\tt
  hep-th/0403141}}].

\bibitem{Bazeia2006f}
D.~Bazeia, F.~A. Brito, and L.~Losano, {\it Scalar fields, bent branes, and
  \text{RG} flow},  {\em J. High Energy Phys.} {\bf 11} (2006) 064,
  [\href{http://xxx.lanl.gov/abs/hep-th/0610233}{{\tt hep-th/0610233}}].

\bibitem{Dzhunushaliev2006}
V.~Dzhunushaliev, H.-J. Schmidt, K.~Myrzakulov, and R.~Myrzakulov, ``Thick
  brane solution with two scalar fields.'' 2006.

\bibitem{Dzhunushaliev200713}
V.~Dzhunushaliev, {\it Thick brane solution in the presence of two interacting
  scalar fields},  {\em Grav. Cosmol.} {\bf 13} (2007) 302,
  [\href{http://xxx.lanl.gov/abs/gr-qc/0603020}{{\tt gr-qc/0603020}}].

\bibitem{George2007}
D.~P. George and R.~R. Volkas, {\it Kink modes and effective four dimensional
  fermion and higgs brane models},  {\em Phys. Rev. D} {\bf 75} (2007) 105007.

\bibitem{Dzhunushaliev200877}
V.~Dzhunushaliev, V.~Folomeev, D.~Singleton, and S.~Aguilar-Rudametkin, {\it 6d
  thick branes from interacting scalar fields},  {\em Phys. Rev. D} {\bf 77}
  (2008) 044006, [\href{http://xxx.lanl.gov/abs/hep-th/0703043}{{\tt
  hep-th/0703043}}].

\bibitem{Bazeia2009c}
D.~Bazeia, A.~R. Gomes, L.~Losano, and R.~Menezes, {\it Braneworld models of
  scalar fields with generalized dynamics},  {\em Phys. Lett. B} {\bf 671}
  (2009) 402, [\href{http://xxx.lanl.gov/abs/0808.1815}{{\tt
  arXiv:0808.1815}}].

\bibitem{Burnier2009}
Y.~Burnier and K.~Zuleta, {\it Effective action of a five-dimensional domain
  wall},  {\em J. High Energy Phys.} {\bf 0905} (2009) 065,
  [\href{http://xxx.lanl.gov/abs/0812.2227}{{\tt arXiv:0812.2227}}].

\bibitem{Chumbes2010}
A.~E.~R. Chumbes and M.~B. Hott, {\it Non-polynomial potentials with deformable
  topological structures},  {\em Phys. Rev. D} {\bf 81} (2010) 045008,
  [\href{http://xxx.lanl.gov/abs/0905.4715}{{\tt arXiv:0905.4715}}].

\bibitem{Dzhunushaliev200979}
V.~Dzhunushaliev, V.~Folomeev, and M.~Minamitsuji, {\it Thick de sitter brane
  solutions in higher dimensions},  {\em Phys. Rev. D} {\bf 79} (2009) 024001,
  [\href{http://xxx.lanl.gov/abs/0809.4076}{{\tt arXiv:0809.4076}}].

\bibitem{Liu2009c}
Y.-X. Liu, Y.~Zhong, and K.~Yang, {\it Scalar-kinetic branes},  {\em Europhys.
  Lett.} {\bf 90} (2010) 51001, [\href{http://xxx.lanl.gov/abs/0907.1952}{{\tt
  arXiv:0907.1952}}].

\bibitem{Dolan1974}
L.~Dolan and R.~Jackiw, {\it Symmetry behavior at finite temperature},  {\em
  Phys. Rev. D} {\bf 9} (1974) 3320.

\bibitem{Jackiw1975}
R.~Jackiw, {\it Symmetry restoration at finite temperature},  {\em Lecture
  Notes in Physics} {\bf 39} (1975) 319.

\bibitem{Weinberg1974}
S.~Weinberg, {\it Gauge and global symmetries at high temperature},  {\em Phys.
  Rev. D} {\bf 9} (Jun, 1974) 3357.

\bibitem{Ansari200722}
R.~U.~H. Ansari and P.~K. Suresh, {\it On the one-loop correction of $\phi^4$
  theory in higher dimensions},  {\em Int. J. Mod. Phys. A} {\bf 22} (2007)
  5369, [\href{http://xxx.lanl.gov/abs/hep-th/0603072}{{\tt hep-th/0603072}}].

\bibitem{KarchRandall2001}
A.~Karch and L.~Randall, {\it Locally localized gravity},  {\em J. High Energy
  Phys.} {\bf 05} (2001) 008,
  [\href{http://xxx.lanl.gov/abs/hep-th/0011156}{{\tt hep-th/0011156}}].

\bibitem{Gremm2000}
M.~Gremm, {\it Thick domain walls and singular spaces},  {\em Phys. Rev. D}
  {\bf 62} (2000) 044017, [\href{http://xxx.lanl.gov/abs/hep-th/0002040}{{\tt
  hep-th/0002040}}].

\bibitem{Duan1958}
I.~S. Duan {\em J. Exp. Theor. Phys.} {\bf 34} (1958) 632.

\bibitem{Fischbach1981}
E.~Fischbach, B.~S. Freeman, and W.-K. Cheng, {\it General-relativistic effects
  in hydrogenic systems},  {\em Phys. Rev. D} {\bf 23} (1981) 2157.

\bibitem{Zhao200776}
Z.-H. Zhao, Y.-X. Liu, and X.-G. Lee, {\it Energy-level shifts of a stationary
  hydrogen atom in a static external gravitational field with schwarzschild
  geometry},  {\em Phys. Rev. D} {\bf 76} (2007) 064016.

\bibitem{Liu2009f}
Y.-X. Liu, H.-T. Li, Z.-H. Zhao, J.-X. Li, and J.-R. Ren, {\it Fermion
  resonances on multi-field thick branes},  {\em J. High Energy Phys.} {\bf 10}
  (2009) 091, [\href{http://xxx.lanl.gov/abs/0909.2312}{{\tt
  arXiv:0909.2312}}].

\bibitem{GonzalezaThompsonb1997}
J.~Gonz{\'a}leza and D.~Thompsonb, {\it Getting started with numerov’s
  method},  {\em Computers in Physics} {\bf 11} (1997), no.~5 514.

\bibitem{Almeida2009}
C.~A.~S. Almeida, R.~Casana, \text{M. M. Ferreira, Jr.}, and A.~R. Gomes, {\it
  Fermion localization and resonances on two-field thick branes},  {\em Phys.
  Rev. D} {\bf 79} (2009) 125022,
  [\href{http://xxx.lanl.gov/abs/0901.3543}{{\tt arXiv:0901.3543}}].

\bibitem{Liu2009}
Y.-X. Liu, J.~Yang, Z.-H. Zhao, C.-E. Fu, and Y.-S. Duan, {\it Fermion
  localization and resonances on a de sitter thick brane},  {\em Phys. Rev. D}
  {\bf 80} (2009) 065019, [\href{http://xxx.lanl.gov/abs/0904.1785}{{\tt
  arXiv:0904.1785}}].

\bibitem{Cruz2009}
W.~T. Cruz, A.~R. Gomes, and C.~A.~S. Almeida, ``Resonances on deformed thick
  branes.'' 2009.

\bibitem{GregoryRubakovSibiryakov2000a}
R.~Gregory, V.~A. Rubakov, and S.~M. Sibiryakov, {\it Opening up extra
  dimensions at ultra-large scales},  {\em Phys. Rev. Lett.} {\bf 84} (2000)
  5928, [\href{http://xxx.lanl.gov/abs/hep-th/0002072}{{\tt hep-th/0002072}}].

\bibitem{FackBerghe1987}
V.~Fack and G.~V. Berghe, {\it (extended) numerov method for computing
  eigenvalues of specific schrodinger equations},  {\em J. Phys. A: Math. Gen.}
  {\bf 20} (1987), no.~13 4153.

\bibitem{CsakiErlichHollowood2000}
C.~Cs\'{a}ki, J.~Erlich, and T.~J. Hollowood, {\it Quasi-localization of
  gravity by resonant modes},  {\em Phys. Rev. Lett.} {\bf 84} (2000) 5932,
  [\href{http://xxx.lanl.gov/abs/hep-th/0002161}{{\tt hep-th/0002161}}].

\bibitem{DvaliGabadadzePorrati2000}
G.~R. Dvali, G.~Gabadadze, and M.~Porrati, {\it Metastable gravitons and
  infinite volume extra dimensions},  {\em Phys. Lett. B} {\bf 484} (2000) 112,
  [\href{http://xxx.lanl.gov/abs/hep-th/0002190}{{\tt hep-th/0002190}}].

\bibitem{GregoryRubakovSibiryakov2000}
R.~Gregory, V.~A. Rubakov, and S.~M. Sibiryakov, {\it Brane worlds: The gravity
  of escaping matter},  {\em Class. Quant. Grav.} {\bf 17} (2000) 4437,
  [\href{http://xxx.lanl.gov/abs/hep-th/0003109}{{\tt hep-th/0003109}}].

\end{thebibliography}\endgroup
\end{document}